\documentclass[aps,floatfix,showpacs,superscriptaddress,showkeys]{revtex4}

\usepackage{amssymb,amsmath,amsfonts,latexsym,graphicx,epsfig}
\usepackage{color} 
    
\usepackage{titlesec}
\titleformat{\section}{\large\bfseries}{\thesection}{1em}{}

\newcommand{\bea}{\begin{eqnarray}}
\newcommand{\ena}{\end{eqnarray}}
\newcommand{\nn}{\nonumber\\}
\newcommand{\be}{\begin{equation}}
\newcommand{\en}{\end{equation}}

\newcommand{\ed}{\end{document}}

\newcommand{\slp}{p\kern-5pt/}
\newcommand{\Tr}{\mbox{\rm{tr}}}

\begin{document}
                
\hfill DSF-2015-1 (Napoli), MITP/15-008 (Mainz) 

\title{Semileptonic decay 
$\Lambda_b \to \Lambda_c+\tau^{-}+\bar \nu_{\tau}$  \\
in the covariant confined quark model} 

\author{Thomas Gutsche}\affiliation{
Institut f\"ur Theoretische Physik, Universit\"at T\"ubingen,
Kepler Center for Astro and Particle Physics, 
Auf der Morgenstelle 14, D-72076, T\"ubingen, Germany}

\author{Mikhail A. Ivanov}
\affiliation{Bogoliubov Laboratory of Theoretical Physics, 
Joint Institute for Nuclear Research, 141980 Dubna, Russia}

\author{J\"{u}rgen G. K\"{o}rner}
\affiliation{PRISMA Cluster of Excellence, Institut f\"{u}r Physik, 
Johannes Gutenberg-Universit\"{a}t, 
D-55099 Mainz, Germany}

\author{Valery~E.~Lyubovitskij}
\affiliation{
Institut f\"ur Theoretische Physik, Universit\"at T\"ubingen,
Kepler Center for Astro and Particle Physics, 
Auf der Morgenstelle 14, D-72076, T\"ubingen, Germany}
\affiliation{ 
Department of Physics, Tomsk State University,  
634050 Tomsk, Russia} 
\affiliation{Mathematical Physics Department, 
Tomsk Polytechnic University, 
Lenin Avenue 30, 634050 Tomsk, Russia} 

\author{Pietro Santorelli}
\affiliation{
Dipartimento di Fisica, Universit\`a di Napoli
Federico II, Complesso Universitario di Monte Sant' Angelo,
Via Cintia, Edificio 6, 80126 Napoli, Italy} 
\affiliation{
Istituto Nazionale di Fisica Nucleare, Sezione di Napoli, 
80126 Napoli, Italy} 

\author{Nurgul Habyl}
\affiliation{Department of Physics and Technology,
Al-Farabi Kazakh National University, 
480012 Almaty, Kazakhstan}
\today

\begin{abstract}
Recently there has been much interest in the tauonic semileptonic meson decays
$B \to D+\tau +\nu_{\tau}$ and  $B \to D^{\ast}+\tau +\nu_{\tau}$, where 
one has found larger rates than what is predicted by the Standard Model. 
We analyze the corresponding semileptonic baryon decays 
$\Lambda^0_b \to \Lambda^+_c + \tau^{-} +\bar \nu_{\tau}$
with particular emphasis on the lepton helicity flip and scalar contributions 
which vanish for zero lepton masses. We calculate the total rate,
the differential decay distributions, the longitudinal and transverse
polarization of the daughter baryon $\Lambda^+_c$ and the $\tau$ lepton, 
and the lepton-side forward-backward asymmetries. 
The nonvanishing polarization of the daughter baryon
$\Lambda^+_c$ leads to hadron-side asymmetries in e.g. the decay
$\Lambda^+_c \to \Lambda^0 + \pi^+$ and azimuthal correlations between
the two final-state decay planes which we specify.
We provide numerical results on these observables using results of the 
covariant confined quark model. We find large lepton mass effects in the
$q^{2}$ spectra and in the polarization 
observables. 
\end{abstract}


\maketitle

\newpage

\section{Introduction}

Recently there has been much discussion about tensions and discrepancies 
in some of the experimental results on leptonic, 
semileptonic and rare decays involving $\mu$ and $\tau$ leptons 
with the predictions of the Standard Model (SM).
Among these are the tauonic $B$ decays 
$\bar B\to \tau \bar\nu_\tau$, $\bar B\to D\,\tau\bar\nu_\tau$ and
$\bar B\to D^\ast\,\tau\bar\nu_\tau$ and the muonic decays 
$B \to K^{\ast}\mu^{+}\mu^{-}$
and $\mathrm{Br}[B \to K\mu^{+}\mu^{-}]/
     \mathrm{Br}[B \to Ke^{+}e^{-}]$. The situation has
been nicely summarized in Refs.~\cite{Soffer:2014kxa,Celis:2014cva,%
Crivellin:2014kga,Blake:2015tda}.
The biggest discrepancies with SM predictions have been reported 
by the {\it BaBar}~\cite{Lees:2012xj} and 
the Belle collaborations~\cite{Matyja:2007kt,Bozek:2010xy}
for the decays $B\to D^{(\ast)}\,\tau\bar\nu_\tau$.  
The discrepancy with the SM results has been summarized in 
Ref.~\cite{Sakaki:2014sea} by comparing
the SM predictions for the ratios
\be
R\left( {\cal D}\right) \equiv
\frac{ \mathrm{Br}\left( \bar B\to  {\cal D}\,\tau^-\bar\nu_\tau \right) }
     { \mathrm{Br}\left( \bar B\to  {\cal D}\,\ell^-\bar\nu_\ell \right) }
\qquad({\cal D} = D,D^\ast; \quad  \ell = e,\mu)
\en
with the experimental results combined from {\it Babar} and Belle.  
The SM predictions were found to be smaller than the measurements by
almost $3.5\,\sigma$: 
\bea
R(D) = \left\{ \begin{array}{l} 
                     0.305\pm 0.012 \qquad \text{SM} \\[1.2ex]
                     0.421\pm 0.058 \qquad \text{{\it Babar} \& Belle}\, 
                 \end{array} \right.
\ena
\bea
R(D^\ast) = \left\{ \begin{array}{l} 
                     0.252\pm 0.004 \qquad \text{SM} \\[1.2ex]
                     0.337\pm 0.025 \qquad \text{{\it Babar} \& Belle}\,. 
                 \end{array} \right.
\ena 
This observation has 
inspired a number of searches for new physics beyond the SM (BSM) in
charged current interactions. These include the addition of
new effective vector, scalar and  tensor interactions in addition to
the standard $V-A$  interaction. Details can be
found in the recent literature on this subject (see, e.g. 
Refs.~\cite{Nierste:2008qe,Pich:2009sp,Jung:2010ik,Faller:2011nj,%
Celis:2012dk,Fajfer:2012jt,Fajfer:2012vx,Datta:2012qk,Crivellin:2012ye,%
He:2012zp,Tanaka:2012nw,Biancofiore:2013ki,Duraisamy:2014sna,Sakaki:2014sea}).

Motivated by the discrepancy between theory and experiment in the meson 
sector, we analyze the corresponding semileptonic baryon decays 
$\Lambda_b^0 \to \Lambda^+_c + \tau^{-} +\bar \nu_{\tau}$
within the SM with particular emphasis on the lepton helicity flip 
contributions which vanish for zero lepton masses. We collect the 
necessary tools to analyze the semileptonic decay 
$\Lambda_b^0 \to \Lambda_c^{+} + \ell^{-} + \bar \nu_{\ell}$ 
$(\ell=e,\mu,\tau)$
as well as the corresponding cascade decay 
$\Lambda_b^0 \to \Lambda^{+}_c 
(\to \Lambda^0 + \pi^+) + \ell^{-} + \bar \nu_{\ell}$.  
As in Refs.~\cite{Korner:1987kd,Korner:1989ve,Korner:1989qb}, we describe the 
semileptonic decays using the helicity formalism which allows one to include 
lepton mass and polarization effects without much additional effort. 

We calculate the total rate, the differential decay distributions, 
the longitudinal and transverse polarization of the daughter baryon 
and lepton-side forward-backward asymmetries. 
The nonvanishing polarization of the daughter baryon
$\Lambda^+_c$ leads to hadron-side asymmetries in e.g. the decay
$\Lambda^+_c \to \Lambda + \pi^+$ and azimuthal correlations between
the two final.state decay planes which we specify.
We provide numerical results on these observables using results of the 
covariant confined quark model. 

We have split the paper into a model-independent and a model-dependent
part. In the first part of the paper, we set up 
the model-independent helicity analysis leading to compact expressions 
for angular decay distributions and polarization observables. In the second
part, we discuss the dynamics of the current--induced 
$\Lambda_{b} \to \Lambda_{c}$ transitions in terms of the covariant
confined quark model and present our numerical results. 

The paper is organized as follows: In Sec.~II, we briefly review
the helicity formalism for the $\Lambda_{b} \to \Lambda_{c}$ transitions and
write down the relations between the helicity 
and invariant transition amplitudes. In Sec.~III, we derive a two-fold
angular decay distribution for the three--body decay process 
$\Lambda^0_b \to \Lambda^+_c + \ell^{-} +\bar \nu_{\ell}$ ($\ell=e,\mu,\tau$).
We define a lepton-side forward-backward asymmetry as well as a convexity 
parameter which describes the $\cos\theta$ dependence of the angular decay
distribution. In Sec.~IV, we determine the longitudinal and 
transverse polarization components of the daughter
baryon $\Lambda^{+}_{c}$ and the charged lepton $\ell^{-}$.  
In Sec.~V, we present the full fourfold
angular decay distribution for the cascade decay
$\Lambda^0_b \to \Lambda^+_c(\to \Lambda^0 + \pi^+) 
+ \ell^{-} +\bar \nu_{\ell}$, where we use the narrow-width approximation
for the intermediate baryon state $\Lambda^+_c$.
In Sec.~VI, we present our dynamical input in terms of the covariant
confined quark model previously developed by us.
We specify the form of the interpolating three-quark currents for the 
$\Lambda_Q$ baryons which are needed in our calculation. Our numerical
results for the invariant $\Lambda_{b} \to \Lambda_{c}$ transition form 
factors are presented in terms of simple double--pole rational approximants.
In Sec.~VII, we present our numerical results for the decay distributions
and polarization observables defined in Secs.~III--V. Finally, Sec.~VIII
contains our summary and conclusions. We have collected some technical 
material in the appendixes. In Appendix~A, we use covariant methods to 
calculate the helicity
components of the unpolarized and polarized lepton tensors of the decays 
$W^{-}_{\rm off-shell}\to \ell^{-}\bar \nu_{\ell}$
and $W^{+}_{\rm off-shell}\to \ell^{+} \nu_{\ell}$ including lepton mass 
effects. In Appendix~B, we present results on the helicity amplitudes for the
above two leptonic transitions.

\section{Invariant and helicity amplitudes}

The matrix element of the process 
$\Lambda_b^0(p_1)\to \Lambda_c^+(p_2) + W^-_{\rm off-shell}(q)$ 
is expressed via the vector and axial vector current matrix elements 
which can be expanded in terms of a complete set of invariants: 
\bea
M_\mu^V(\lambda_{1},\lambda_{2}) &=& \langle B_2,\lambda_{2}|J_\mu^V|
B_1,\lambda_{1}\rangle
  = \bar u_2(p_2,\lambda_{2})\bigg[F_1^V(q^2)\gamma_\mu-\frac{F_2^V(q^2)}{M_1}
  i\sigma_{\mu\nu}q^\nu+\frac{F_3^V(q^2)}{M_1}q_\mu\bigg]u_1(p_1,\lambda_{1}),
\nonumber\\
\label{eq: invariants}\\
M_\mu^A(\lambda_{1},\lambda_{2}) &=& \langle B_2,\lambda_{2}|J_\mu^A|
B_1,\lambda_{1}\rangle
  = \bar u_2(p_2,\lambda_{2})\bigg[F_1^A(q^2)\gamma_\mu-\frac{F_2^A(q^2)}{M_1}
  i\sigma_{\mu\nu}q^\nu+\frac{F_3^A(q^2)}{M_1}q_\mu\bigg]\gamma_{5}
u_1(p_1,\lambda_{1}) 
\ena
where $\sigma_{\mu\nu}=\frac i2(\gamma_\mu\gamma_\nu-\gamma_\nu\gamma_\mu)$ 
and  $q = p_1 - p_2$. The labels $\lambda_{i}=\pm \frac12$ denote the 
helicities of the two baryons. In the present application $B_1=\Lambda_b$ and 
$B_2=\Lambda_c$.

The SM current is not conserved and thus consists of a superposition of a 
spin-1 and a spin-0 
component where the $J^{P}$ content of the vector current $J^V_{\mu}$ and 
the axial vector current $J^A_{\mu}$ are $(0^{+},1^{-})$ and $(0^{-},1^{+})$,
respectively. One defines helicity amplitudes through
\be
H^{V/A}_{\lambda_2\lambda_W} = 
M_\mu^{V/A}(\lambda_2)\epsilon^{\dagger\,\mu}(\lambda_W)\,, 
\label{eq:hel_def}\,‚
\en
where there are four  helicities for the $W^{-}_{\rm off-shell}$,
namely $\lambda_W=\pm 1,0\,(J=1)$ and $\lambda_W=0\,(J=0)$. The label $J=1,0$ 
denotes the two angular momenta of the rest frame $W^{-}_{\rm off-shell}$. 
In order to distinguish the two $\lambda_W=0$ states we follow 
the convention of Refs.~\cite{Korner:1989ve,Korner:1989qb} and adopt
the notation $\lambda_W=0$ for $J=1$ and $\lambda_W=t$ for $J=0$ 
($t$ for temporal). From angular momentum conservation, one has 
$\lambda_{1}=-\lambda_{2}+\lambda_{W}$. 

It is easiest to calculate the helicity amplitudes in the rest frame of the 
parent baryon $B_{1}$, where we choose the $z$--axis to be along the 
$W^{-}_{\rm off-shell}$ (see Fig.~\ref{fig:angles}). They read 
(see e.g. Refs.~\cite{Gutsche:2013pp,Gutsche:2013oea,Gutsche:2014zna})

\bea
H_{+\frac12 t}^{V/A}&=&\frac{\sqrt{Q_\pm}}{\sqrt{q^2}}
  \bigg( M_\mp F_1^{V/A}\pm \frac{q^2}{M_1} F_3^{V/A}\bigg),
\nn
H_{+\frac12 +1}^{V/A}&=&\sqrt{2Q_\mp}
  \bigg(F_1^{V/A}\pm \frac{M_\pm}{M_1}F_2^{V/A}\bigg),
\nn
H_{+\frac12 0}^{V/A}&=&\frac{\sqrt{Q_\mp}}{\sqrt{q^2}}
  \bigg(M_\pm F_1^{V/A}\pm \frac{q^2}{M_1} F_2^{V/A}\bigg)\,. 
\label{eq:hel_inv}
\ena
where we make use of the abbreviations $M_\pm = M_1\pm M_2$ and 
$Q_\pm = M_\pm^2  - q^2$.
 
From parity or from an explicit calculation, one has
\bea
H_{-\lambda_2,-\lambda_W}^V  = H_{\lambda_2,\lambda_W}^V,
\qquad
H_{-\lambda_2,-\lambda_W}^A = -H_{\lambda_2,\lambda_W}^A.
\ena 
The total left--chiral helicity amplitude is defined by the composition
\bea
H_{\lambda_2,\lambda_W} = H_{\lambda_2,\lambda_W}^V 
- H_{\lambda_2,\lambda_W}^A .
\ena 

The polarization observables to be discussed further on can be expressed
in terms of helicity structure functions given in terms of  bilinear 
combinations of helicity amplitudes. The definitions of the structure
functions are collected
in Table~\ref{tab:bilinears}. Eleven of these contribute to the partial
rates and polarization 
observables discussed in this paper. For
the sake of completeness, we have also listed the structure function
${\cal H}_{ST}$, which enters in the description of polarized 
$\Lambda_{b}$ decay which we, however, do not discuss in this paper.
\begin{table}[ht] 
\begin{center}
\caption{ Definition of helicity structure functions and their parity 
properties.}
\def\arraystretch{2}
\begin{tabular}{ll}
\hline
Parity-conserving (p.c.) \qquad & \qquad  parity-violating (p.v.)  \\
\hline
${\cal H}_U   = |H_{+\frac12 +1}|^2 + |H_{-\frac12 -1}|^2$   \qquad &  \qquad
${\cal H}_P   = |H_{+\frac12 +1}|^2 - |H_{-\frac12 -1}|^2$   \\
${\cal H}_L    = |H_{+\frac12\, 0}|^2 + |H_{-\frac12\, 0}|^2$   \qquad &  \qquad
${\cal H}_{L_P} = |H_{+\frac12\, 0}|^2 - |H_{-\frac12\, 0}|^2$  \\
${\cal H}_S    = |H_{+\frac12\, t}|^2 + |H_{-\frac12\, t}|^2$   \qquad &  \qquad
${\cal H}_{S_P} = |H_{+\frac12\, t}|^2 - |H_{-\frac12\, t}|^2$   \\
${\cal H}_{LT}  =   {\rm Re}\left(  H_{+\frac12 +1}  H_{-\frac12\, 0}^\dagger 
                         + H_{+\frac12\, 0} H_{-\frac12 -1}^\dagger \right)$ 
 \qquad &  \qquad
${\cal H}_{LT_P} =   {\rm Re}\left(  H_{+\frac12 +1}  H_{-\frac12\, 0}^\dagger 
                           - H_{+\frac12\, 0} H_{-\frac12 -1}^\dagger \right)$ \\
 ${\cal H}_{ST} =  {\rm Re}\left( H_{+\frac12 +1} H_{-\frac12\,t}^\dagger 
                        + H_{+\frac12\,t} H_{-\frac12 -1}^\dagger \right)$
 \qquad &  \qquad
${\cal H}_{ST_P} =  {\rm Re}\left(  H_{+\frac12 +1} H_{-\frac12\,t}^\dagger 
                          - H_{+\frac12\,t} H_{-\frac12 -1}^\dagger \right)$  \\
${\cal H}_{SL} =  {\rm Re}\left(  H_{+\frac12\,0} H_{+\frac12\,t}^\dagger 
                        + H_{-\frac12\,0} H_{-\frac12\,t}^\dagger \right) $
 \qquad &  \qquad
${\cal H}_{SL_P} =  {\rm Re}\left(  H_{+\frac12\,0} H_{+\frac12\,t}^\dagger 
                         - H_{-\frac12\,0} H_{-\frac12\,t}^\dagger \right)$ \\[2ex]
\hline
\end{tabular}
\label{tab:bilinears}
\end{center}
\end{table}

The helicity structure functions have definite parity properties as 
indicated in Table \ref{tab:bilinears}. The first and second columns of 
Table~\ref{tab:bilinears} list the parity-conserving (p.c.)
and parity-violating (p.v.) bilinear combinations of helicity amplitudes,
respectively; i.e., the p.c. and p.v. helicity structure functions result from
products of $VV$ and $AA$, and $VA$ and $AV$ currents, respectively.

The vector and axial vector helicity amplitudes possess simple structures
in the kinematical limits of zero and maximal recoil. At the zero recoil 
point $q^{2} \to (M_{1}-M_{2})^{2}$, one has only $s$--wave transitions 
(conventionally called allowed Fermi and allowed 
Gamow-Teller transitions, respectively). 
The surviving helicity amplitudes are
\bea
H^{V}_{\frac12 t}&=&H^{V}_{-\frac12 t}=2\sqrt{M_{1}M_{2}}\,(F_{1}^{V}+ 
\frac{M_{-}}{M_{1}}F_{3}^{V})\qquad {\rm allowed \,\,Fermi\,,}
\label{eq:fermi}\\
H^{A}_{\frac12 1}/\sqrt{2}&=&H^{A}_{\frac12 0}=2\,\sqrt{M_{1}M_{2}}\,
(F_{1}^{A} -
\frac{M_{-}}{M_{1}}F_{2}^{A})\qquad{\rm allowed \,\,Gamow-Teller\,.}
\label{eq:gamow}
\ena

The zero recoil structure has implications for the structure functions.
At zero recoil, one finds
\be
{\cal H}_{ST},\,{\cal H}_{SL},\,{\cal H}_{P},\,{\cal H}_{L_{P}},
\,{\cal H}_{S_{P}},{\cal H}_{LT_{P}}=0\,, 
\qquad {\cal H}_{U}=2\,{\cal H}_{L}=\sqrt{2}\,{\cal H}_{LT}.
\label{eq:zerec}
\en
At $q^{2}=0$ (which is very close to the maximal recoil point for the 
$e$ and $\mu$ modes), the scalar and longitudinal helicity 
amplitudes dominate. At $q^{2}=0$ one has
\be
H_{\pm\frac12 t}=H_{\pm\frac12 0}\sim(F_{1}^{V}\mp F_{1}^{A}).
\label{eq:q2zero}
\en

As we shall see later on, the results of our dynamical quark model
are quite close to the heavy quark effective theory 
HQET result $F_{1}^{V}(q^{2})=F_{1}^{A}(q^{2})=F(q^{2}),\,F_{2,3}^{V/A}=0$
(see e.g. the review in Ref.~\cite{Korner:1994nh}).
It is therefore quite useful to list the HQET relations for the helicity
amplitudes and structure functions. In the HQET limit, one has
\be
H_{\frac12 0}=-\,H_{\frac12 t}\,,  \qquad
H_{-\frac12 0}=\,H_{-\frac12 t}\,, 
\label{hqet1}
\en
which leads to the structure function relations
\be
{\cal H}_{S_{P}}={\cal H}_{L_{P}}=-{\cal H}_{SL}\,, \qquad 
{\cal H}_{L}={\cal H}_{S}=-{\cal H}_{SL_{P}}\,,     \qquad 
{\cal H}_{ST}={\cal H}_{LT_{P}}\,,                  \qquad 
{\cal H}_{ST_{P}}={\cal H}_{LT}.
\label{hqet2}
\en
If one combines the zero recoil structure with the HQET results, one finds
the zero recoil amplitude relations
\be
H_{\frac12 t}=H_{\frac12 0}=H_{\frac12 1}/\sqrt{2}=H_{-\frac12 t}=
-H_{\frac12 0}
=-H_{-\frac12 -1}/\sqrt{2} .
\label{hqet3}
\en 
At maximal recoil for the ($e,\mu$) modes the HQET 
relations~(\ref{eq:q2zero}) turn into the helicity amplitude relations 
\be
H_{\frac12 t}=H_{\frac12 0}=0\,, \qquad 
H_{-\frac12 t}=H_{-\frac12 0} \,. 
\label{hqet4}
\en

We have gone to great lengths to describe the limiting values of the
structure functions at both ends of the kinematical $q^{2}$ range.
These limiting values allow us to understand most of the the limiting
behavior of our numerical results on the
polarization observables at zero recoil and maximal recoil to be discussed
in Sec.~VII. Similar limiting relations have been discussed before 
in Ref.~\cite{Korner:1991ph} for the zero lepton mass case.

The helicity amplitudes $H_{\lambda_2,\lambda_W}$ are a superposition of
vector and axial vector pieces and thus 
do not have definite parity properties. One can project back to the
vector and axial vector helicity amplitudes by defining the transversity
amplitudes~\cite{Kruger:2005ep}: 
\bea
    A_{\lambda_{2}\parallel/\perp}
&=&(H^{V-A}_{\lambda_{2}+1}
\pm H^{V-A}_{-\lambda_{2}-1})/\sqrt{2}, 
\nn
    A^{V/A}_{\lambda_{2}0}&=&(H^{V-A}_{\lambda_{2}0}
\pm H^{V-A}_{-\lambda_{2}0})/\sqrt{2} ,
\nn
A^{V/A}_{\lambda_{2}t}&=&(H^{V-A}_{\lambda_{2}t}
\pm H^{V-A}_{-\lambda_{2}t})/\sqrt{2}.
\label{eq:parity1}
\ena
From Eq.~(\ref{eq:parity1}), one can see that the 
transversity amplitudes have definite 
transformation properties under parity, e.g. $A_{\lambda_{2}
\parallel}=A_{-\lambda_{2}\parallel}$,\,
$A_{\lambda_{2}
\perp}=-A_{-\lambda_{2}\perp}$, etc. 

For the structure functions in Table~\ref{tab:bilinears}, one obtains
e.g.
\be {\cal H}_{U}=|A_{\lambda_{2}\parallel}|^{2}+|A_{\lambda_{2}\perp}|^{2}\,, 
\qquad \qquad {\cal H}_{P}=2\,{\rm Re}\,
\big(A_{\lambda_{2}\parallel}A_{\lambda_{2}\perp}^{\dagger}\big)\,.
\en 
The transversity amplitudes correspond to the Cartesian 
components of the helicity amplitudes. The parity properties of the
various structure functions in the transversity representation are 
clearly manifest. They are identical to those in the helicity representation
where they are not quite manifest.

\section{Twofold angular decay distribution}

We first consider the three-particle decay 
$\Lambda_b^0(p_1)\to \Lambda_c^+(p_2) 
+ W^-_{\rm off-shell}\Big( \to \ell^-(p_\ell) 
+ \bar\nu_\ell(p_{\nu_\ell})\Big)$ 
with $q=p_\ell+p_{\nu_\ell}$, $p_\ell^2 = m_\ell^2$, and $p_{\nu_\ell}^2 = 0$. 
At the present stage we sum over the helicities of the parent and daughter 
baryon. 
The three-body decay can be described in terms of the invariant 
variable $q^2$ and the polar angle $\theta$ defined in Fig.~\ref{fig:angles}. 
The differential $(q^2,\cos\theta)$ distribution reads
\be
\frac{d\Gamma}{dq^2 d\cos\theta} =  \frac{G^2_F}{(2\pi)^4}|V_{bc}|^2 
\frac{(q^2-m_\ell^2)\, |\mathbf{p_2}|}{128 M_1^2q^2}\, H^{\mu\nu}
\,L_{\mu\nu}(\theta)
\label{eq:distr2}
\en
where $|\mathbf{p_2}| = \lambda^{1/2}(M_1^2,M_2^2,q^2)/(2M_1)=
\sqrt{Q_{+}Q_{-}}/(2M_{1})$ is the momentum
of the daughter baryon in the $\Lambda_b$ rest frame.
The lepton tensor $L_{\mu\nu}$ can be calculated as (see Appendix A) 

\be
L^{\mu \nu}=8\,\biggl( p_\ell^\mu p_{\nu_\ell}^\nu 
+ p_\ell^\nu p_{\nu_\ell}^\mu 
         - \frac12 (q^2-m_\ell^2) g^{\mu\nu}
+ i \varepsilon^{\mu \nu \alpha \beta} 
p_{\ell\alpha} \, p_{\nu_\ell\beta} \biggr) \,,  
\label{eq:lept_tensor1}
\en
where $\varepsilon_{0123}=+1$.
The hadron tensor $H_{\mu\nu}$ is given by the tensor product of the 
chiral matrix elements
\be
H_{\mu\nu} =  M_\mu\,M_\nu^\dagger \,, 
\qquad M_\mu = M_\mu^V - M_\mu^A\,,    
\label{eq:hadr_tensor}
\en
where the vector and axial vector matrix elements are
defined in Eq.~\eqref{eq: invariants}. 
 
The $\cos\theta$ dependence of $H_{\mu\nu}L^{\mu\nu}(\theta)$ can be worked
out by following the methods described 
in Ref.~\cite{Korner:1987kd,Korner:1989ve,Korner:1989qb} 
and later on adopted by
Fajfer {\it et al.}~\cite{Fajfer:2012vx} 
in their analysis of BSM effects in the 
decays $\bar B\to D^{(\ast)}\,\tau\bar\nu_\tau$.  
Note that the $\cos\theta$ dependence
can also be worked out by using traditional methods by a straightforward
contraction of $H_{\mu\nu}L^{\mu\nu}(\theta)$. This leads to expressions 
involving 
scalar products of momenta that are defined in different reference frames. 
When doing the required contractions, the four-momenta have to be boosted to 
a common reference frame as e.g. described in Ref.~\cite{Buchalla:2013mpa} for
the decay $H \to Z^0\ell^+\ell^-$ and in Ref.~\cite{Gevorkyan:2014waa} 
for the decay $K^{\pm}\to \pi^{\pm}\pi^0e^+e^-$. 
The advantage of the helicity method is that the origin
of the angular factors multiplying the helicity structure functions can
be straightforwardly identified. One makes use of 
the completeness relation for the polarization four--vectors
\be
 \sum_{m,m'=t,\pm,0} 
\epsilon^\mu(m) \epsilon^{\dagger\, \nu }(m') g_{m m'} 
= g^{\mu \nu} \, .
\label{eq:completeness}
\en 
The tensor $ g_{mm'} = \mbox{diag} (+,-,-,-) $ is the spherical representation
of the metric tensor, where the components are ordered in the sequence 
$m,m'=t,+,0,-$.  

Using the completeness relation~(\ref{eq:completeness}), one can rewrite the 
contraction of the hadron and lepton tensors $H_{\mu\nu}L^{\mu\nu}(\theta)$ 
as a sum over products of their respective helicity components. 
Appropriate to the problem, we replace the
labels $(m,m')$ with the helicity components 
$(\lambda_{W},\lambda'_{W})$ of the $W^-_{\rm off-shell}$. One finds
\bea
H_{\mu\nu}L^{\mu\nu}(\theta) &=& 
H^{\mu'\nu'}\,g_{\mu'\mu}g_{\nu'\nu}\,L^{\mu\nu}(\theta) 
\nn[2ex]
&=& 
\sum_{\lambda_{W},\lambda'_{W}\lambda''_{W},\lambda'''_{W}} 
H^{\mu'\nu'} \epsilon_{\mu'}(\lambda_{W})
\epsilon_{\mu}^\dagger(\lambda'_{W})\,g_{\lambda_{W}\lambda'_{W}}\,
\epsilon_{\nu'}^\dagger(\lambda''_{W}) \epsilon_{\nu}(\lambda'''_{W})
\,g_{\lambda''_{W}\lambda'''_{W}} \, 
L^{\mu\nu}(\theta) \nn[2ex]
&=&\sum_{\lambda_{W},\lambda'_{W}\lambda''_{W},\lambda'''_{W}}
\bigg( H^{\mu'\nu'} \epsilon_{\mu'}(\lambda_{W}) 
\epsilon_{\nu'}^\dagger(\lambda''_{W}) \bigg) \;
    \bigg(L^{\mu\nu}(\theta) \epsilon_{\mu}^\dagger(\lambda'_{W}) 
\epsilon_{\nu}(\lambda'''_{W}) 
\bigg) 
\,g_{\lambda_{W}\lambda'_{W}} g_{\lambda''_{W}\lambda'''_{W}} 
 \nn[2ex]
&\equiv&
\sum_{\lambda_{W}\lambda'_{W}} H_{\lambda_{W}\lambda'_{W}}
L_{\lambda_{W}\lambda'_{W}}(\theta)g_{\lambda_{W}\lambda_{W}}
g_{\lambda'_{W}\lambda'_{W}}\,.
\label{eq:Lorentz_trick}
\ena

The two factors enclosed by round brackets in 
the third line of Eq.(\ref{eq:Lorentz_trick}) are Lorentz invariant and 
can thus be evaluated in different Lorentz frames. The leptonic part will be 
evaluated in the $(\ell,\nu_\ell)$ CM frame (or $W_{\rm{off-shell}}$ rest 
frame) with the positive $z$ axis along the $W_{\rm off-shell}$ bringing 
in the decay angle $\theta$. The hadronic part is 
evaluated in the $\Lambda_b$ rest frame bringing in the helicity
amplitudes [Eq.~(\ref{eq:hel_inv})].

The product of the two spherical metric tensors
$g_{\lambda_{W}\lambda_{W}}g_{\lambda'_{W}\lambda'_{W}}$ in the last line of 
(\ref{eq:Lorentz_trick}) gives
$+1$ for $\lambda_{W}=t,\,\lambda'_{W}=t$ and 
$\lambda_{W}=i,\,\lambda'_{W}=j\,(i,j=1,0,-1)$ and $-1$ for
$\lambda_{W}=t,\,\lambda'_{W}=i\,(i=1,0,-1)$ and 
$\lambda_{W}=i,\,\lambda'_{W}=t\,(i=1,0,-1)$; i.e., there is an extra 
minus sign for the spin 0--spin 1 interference
contribution. We can therefore rewrite the last line of 
Eq.~(\ref{eq:Lorentz_trick}) as
\be
H_{\mu\nu}L^{\mu\nu}(\theta)=\sum_{\lambda_{W},\lambda'_{W}}
(-1)^{J+J'}H_{\lambda_{W}\lambda'_{W}}L_{\lambda_{W}\lambda'_{W}}(\theta) \,,
\label{eq:HL}
\en
where the sum is over $\lambda_{W},\lambda'_{W}=t,\pm 1,0$. 
Note that, through the choice of the ``t'' notation, the dependence of 
$H_{\lambda_{W}\lambda'_{W}}$ and
$L_{\lambda_{W}\lambda'_{W}}(\theta)$ on $J,J'$ is implicit.

Referring to the factorized form Eq.~(\ref{eq:hadr_tensor}), the helicity
representation of the hadron tensor can also be written 
in terms of bilinear products of the helicity amplitudes, i.e.
\be
{\cal H}_{\lambda_{W}\lambda'_{W}}=\sum_{\lambda_{2}} 
H_{\lambda_{2}\,\lambda_{W}}
H^{\dagger}_{\lambda_{2}\,\lambda'_{W}}\,, 
\label{eq:facform}
\en
where the three helicities of the process satisfy the angular momentum 
constraint $\lambda_{1}=\lambda_{2}-\lambda_{W}$.

One has to remember that Eq.~\eqref{eq:distr2} refers to the 
differential decay rate of an unpolarized parent baryon 
into a daughter baryon whose spin is not observed. Together with the
angular momentum constraint $\lambda_{1}=\lambda_{2}-\lambda_{W}$,  
this implies $\lambda_{W}=\lambda'_{W}$. One therefore has 
diagonal contributions $\lambda_{W}=\lambda'_{W}=t,\pm 1,0$ 
as well as quasi-diagonal contributions with $\lambda_{W}=t$ and 
$\lambda'_{W}=0$ and vice versa.

Using the factorized form~(\ref{eq:facform}), Eq.~(\ref{eq:HL}) turns into
\be
H_{\mu\nu}L^{\mu\nu}(\theta) = 
\sum_{\lambda_W,\lambda'_W,\lambda_2}
(-1)^{J+J'}  
\delta_{\lambda_2 - \lambda_W,\lambda_2 - \lambda'_W}
H_{\lambda_2 \lambda_W}H_{\lambda_2 \lambda'_W}^\dagger 
L_{ \lambda_W  \lambda'_W}(\theta)\,, 
\label{eq:LH2}
\en
where the delta function 
$\delta_{\lambda_2 - \lambda_W,\lambda_2 - \lambda'_W}$ encodes the fact
mentioned above that we are dealing with the decay of an unpolarized
parent baryon $B_{1}$ into a daughter baryon $B_{2}$ whose spin is not 
observed. The helicity components of the lepton tensor 
$L_{ \lambda_W  \lambda'_W}(\theta)$ have been calculated in 
Appendix A~[see Eq.~(\ref{lt2})].

One can go one step further and express the leptonic helicity tensor
$L_{ \lambda_W  \lambda'_W}(\theta)$ in terms of the leptonic helicity
amplitudes $h_{\lambda_{\ell}\,\lambda_{\bar\nu_\ell}=1/2} (J)$ defined
in Appendix B. They describe the decay 
$W^{-}_{\rm{off-shell}}\to \ell^{-}\bar \nu_{\ell}$, where the
$z$--direction is along the charged lepton $\ell^{-}$. We shall refer to 
this frame as the $W$--decay frame. Then one has to rotate the leptonic 
helicity amplitudes defined in the $W$--decay
frame to the original $z$--direction employing Wigner's
$d^{J}_{mm'}(\theta)$ functions (as in Ref.~\cite{Kadeer:2005aq} we use the 
convention of Rose \cite{rose57} for the $d$--functions). One then has 
($d^{0}_{00}=1$) 
\be
L_{ \lambda_W  \lambda'_W}(\theta)= \sum_{\lambda_{\ell}}
d^{J}_{\lambda_W,\lambda_{\ell}-1/2}\,(\theta)\,
\, d^{J'}_{\lambda'_W,\lambda_{\ell}-1/2}\,(\theta)
\,\, h_{\lambda_{\ell}\,1/2}\,(J)
\,\, h^{\dagger}_{\lambda_{\ell}\,1/2}\,(J')\,. 
\label{Lhat}
\en
The leptonic helicity amplitudes $h_{\lambda_{\ell^{-}}\,1/2}(J)$ have been 
calculated in Appendix B. We mention that the representation (\ref{Lhat}) has 
been used to check the results of the covariant determination of 
$L_{ \lambda_W  \lambda'_W}(\theta)$ used in (\ref{eq:LH2}).

Returning to the representation in Eq.~(\ref{eq:LH2}), one finds
using the results of Appendix A ($v=1-m^2_{\ell}/q^2$\,\,)
\be
H_{\mu\nu} L^{\mu\nu}(\theta)=\frac{16 \pi}{3}\,(2q^2v)W(\theta)\,, 
\en 
with
\bea
W(\theta) 
&=& 
      \frac38\,(1 + \cos^{2}\theta)\, {\cal H}_U 
      - \frac34\,\cos\theta\, {\cal H}_P
      + \frac34\,\sin^2\theta\, {\cal H}_L \nonumber\\
&+&\delta_\ell\,\Big\{
        \frac32\, {\cal H}_S 
      + \frac34\,\sin^2\theta\, {\cal H}_U 
      + \frac32\,\cos^2\theta\, {\cal H}_L 
      - 3\cos\theta\,{\cal H}_{SL}\Big\} 
\,.
\label{two1}
\ena
The first three terms in~(\ref{two1}) result from lepton helicity nonflip
($nf$) contributions, while the last four terms proportional to $\delta_{\ell}$
are lepton helicity flip ($hf$) contributions,   
where the helicity flip suppression factor is given by (see Appendix B) 
\be
\label{hflip}
\delta_{\ell}=\frac{m^{2}_{\ell}}{2q^{2}}\,.
\en

It is interesting to note that of the two observables ${\cal H}_{P}$ and 
${\cal H}_{SL}$ that multiply the parity-odd angular factor $\cos\theta$ 
in Eq.~(\ref{two1}), ${\cal H}_{P}$ is parity violating whereas 
${\cal H}_{SL}$ is parity conserving (see Table~\ref{tab:bilinears}). In the 
latter case, the parity-odd nature results from the parity-even but parity-odd
interference of the $(0^{+},1^{-})$ pieces in the $(VV)$ term and the
$(0^{-},1^{+})$ pieces in the $(AA)$ term.

Integrating over $\cos\theta$ and putting in the correct normalization, 
one obtains the normalized differential rate, which reads
\be
\frac{d\Gamma}{dq^2} = 
\Gamma_{0} \frac{(q^2-m_\ell^2)^2 |\mathbf{p_2}|}{M_1^7q^2} \;
\Bigg\{  {\cal H}_U\, +\, {\cal H}_L\, 
+\,  \delta_\ell \, \Big[
 {\cal H}_U\, +\, {\cal H}_L\,+\, 3\,{\cal H}_S
\Big] \;\Bigg\} \equiv 
\Gamma_{0} \frac{(q^2-m_\ell^2)^2 |\mathbf{p_2}|}{M_1^7q^2} \; 
{\cal H}_{\rm tot}\,, 
\label{eq:1-fold}
\en
where 
\be
{\cal H}_{\rm tot}=\int d\cos\theta \,W(\theta)
= {\cal H}_U\, +\, {\cal H}_L\, 
+\,  \delta_\ell \, \Big[
 {\cal H}_U\, +\, {\cal H}_L\,+\, 3\,{\cal H}_S
\Big] \,.
\en
In~(\ref{eq:1-fold}), we have introduced the Born term rate
\be
\Gamma_0=\frac{G_F^2|V_{bc}|^2 M_1^5}{192 \pi^3}\,.
\en
The rate $\Gamma_{0}$ represents the SM rate of the decay of a massive parent
fermion into three massless fermions, i.e. $M_1\neq 0$ and 
$M_2,m_\ell,m_{\nu_\ell}=0$, where $F^{V/A}_{1}=1$ and $F^{V/A}_{2,3}=0$.
The $q^{2}$--dependent factor multiplying $\Gamma_{0}$ in 
Eq.(\ref{eq:1-fold}) can be seen to be integrated to 1 for these mass and form 
factor settings.

It is convenient to define partial rates $d\Gamma_{X}/dq^2$ and
$d\widetilde\Gamma_{X}/dq^2$ for the helicity nonflip (nf) and 
helicity flip (hf) helicity structure functions ${\cal H}_{X}$ defined in 
Table~\ref{tab:bilinears}. $d\Gamma_{X}/dq^2$ refers to a
lepton helicity nonflip ($nf$) rate, while $d\widetilde\Gamma_{X}/dq^2$ 
refers to a lepton helicity flip ($hf$) rate. One has 
\bea
\frac{d\Gamma_{X}}{dq^2}(nf) &=& 
\Gamma_0 \, 
\frac{(q^2-m_\ell^2)^2 |\mathbf{p_2}|}{M_1^7q^2} \, {\cal H}_X\,, 
\quad\quad\quad  X = U,\, L,\, P,\, L_P,\, LT,\, LT_P \,,\nonumber \\
\frac{d\widetilde\Gamma_{X}}{dq^2}(hf) &=& \delta_{\ell}
\,\Gamma_0 \, 
\frac{(q^2-m_\ell^2)^2 |\mathbf{p_2}|}{M_1^7q^2} \, {\cal H}_X\,, 
\quad\quad\quad  X = U,\, L,\,LT,\, P,\, SL,\, S,\, L_P,\, S_P, \,SL_{p},
\, LT_{P},\, ST_P \,.
\ena
The partial rates can then be split into a helicity nonflip ($nf$) and helicity
flip ($hf$) part according to
\be
\frac{d\Gamma_{X}}{dq^2}=\frac{d\Gamma_{X}}{dq^2}(nf)
+\frac{d\widetilde\Gamma_{X}}{dq^2}(hf).
\en
It is clear that the (un-normalized) longitudinal polarization of the charged
lepton for a given helicity structure function ${\cal H}_{X}$ is then given by
\be
P^{\ell}_{z}(X)\sim d\widetilde\Gamma_{X}/dq^{2}-d\Gamma_{X}/dq^{2}
\label{lplep}
\en
The distribution Eq.~(\ref{two1}) is given in terms of a second-order 
polynomial in $\cos\theta$. The linear term results in a forward-backward 
asymmetry defined by
\be
A_{FB}^{\ell}(q^2) =\frac{d\Gamma(F)-d\Gamma(B)}{d\Gamma(F)+d\Gamma(B)}
= -\frac32 \frac{{\cal H}_P\,+4\,\delta_\ell\,{\cal H}_{SL}}{{\cal H}_{\rm tot}} .
\label{eq:FB}
\en
The quadratic term can be isolated by taking the second derivative 
of~(\ref{two1}). We therefore define a convexity parameter $C_F(q^2)$
according to 

\be
C_F(q^2) = \frac{1}{{\cal H}_{\rm tot}} 
\,\frac{d^{2}W(\theta)}{d(\cos\theta)^{2}} 
=\frac34\,(1\,-\,2\delta_\ell)\, 
   \frac{{\cal H}_U\,-\,2\,{\cal H}_{L}}{{\cal H}_{\rm tot}} .
\label{eq:CF}
\en

\section{Polarization of the  daughter baryon $\Lambda_c$ and the charged 
lepton $\ell^-$ 
}
The polarization components of the daughter baryon $B_2$ 
($B_{2}=\Lambda_{c}$ in the present application)
can be obtained from the spin density matrix of the daughter baryon. 
In Eq.~(\ref{eq:LH2}) we have taken the sum over helicities of 
the daughter baryon $B_2$ or, put in a different language, we have taken 
the trace of the polarization density matrix $\rho_{\lambda_2\,\lambda_2'}$ 
of the daughter baryon $B_{2}$. In order to calculate the polarization 
components of the daughter baryon, we need to consider the polarization
density matrix components 
$(\rho_{1/2\, 1/2} - \rho_{-1/2\, -1/2})$ for $P_z$ and 
$2\mathrm{Re}\rho_{1/2\, -1/2}$ for $P_x$; i.e., we have to allow for 
$\lambda_{2} \neq \lambda'_{2}$ in Eq.~(\ref{eq:LH2}). According to
Eq.~(\ref{eq:LH2}) the un-normalized
density matrix components of the daughter baryon $B_{2}$ read
\be
\rho_{\lambda_{2}\,\lambda'_{2}}= 
\sum_{\lambda_W,\lambda'_W}
(-1)^{J+J'}  
\delta_{\lambda_2 - \lambda_W,\lambda'_2 - \lambda'_W}
H_{\lambda_2 \lambda_W}H_{\lambda'_2 \lambda'_W}^\dagger 
L_{ \lambda_W  \lambda'_W}(\theta)\,, 
\en
The polarization can be seen to 
be $\cos\theta$ dependent. Instead of considering the full 
$\theta$--dependent polarization
components we take the mean of $P_x(\theta)$ and $P_z(\theta)$ 
with regard to $\cos\theta$. One obtains 
\bea
P^h_z(q^2)  &=& \frac{\rho_{1/2\, 1/2}-\rho_{-1/2\, -1/2}}
{\rho_{1/2\, 1/2}+\rho_{-1/2\, -1/2}}
=\frac{ {\cal H}_{P} + {\cal H}_{L_P} + \delta_{\ell}\,
({\cal H}_{P} + {\cal H}_{L_P} + 3{\cal H}_{S_P})}{{\cal H}_{\rm tot}}\,, 
 \nn[1.5ex]
P^h_x(q^2) &=&\frac{2\,\mathrm{Re} \,\rho_{1/2\, -1/2}}
{\rho_{1/2\, 1/2}+\rho_{-1/2\, -1/2}}
=-\frac{3\pi}{4\sqrt{2}}\frac{{\cal H}_{LT} 
- 2\,\delta_{\ell}{\cal H}_{ST_P}}{{\cal H}_{\rm tot}}.
\label{eq:Pol_had}
\ena
The corresponding $\theta$--dependent expressions are written down in
Sec.~IV.

In the zero lepton mass limit, the charged lepton $\ell^{-}$ is 
100\% polarized opposite to its momentum direction 
(i.e. it has negative helicity). 
Lepton mass effects bring in helicity flip contributions which can 
considerably change the magnitude of the polarization $|\vec {P}^{\ell}|$ 
and its orientation. In Appendix A we have calculated
the lepton's polarization components $P_x^{\ell}(\theta)$ and 
$P_z^{\ell}(\theta)$.
Again, we consider only the $\cos\theta$ averaged polarization. 
Using the results of Appendix~A, one finds 
\bea
P^\ell_z(q^2)  &=& 
-\frac{{\cal H}_U +{\cal H}_L-\delta_{\ell}\,( {\cal H}_U + {\cal H}_L 
+3{\cal H}_{S} )}{{\cal H}_{\rm tot}}\,, 
\nonumber \\[1.5ex]
P^\ell_x(q^2)  &=& -\frac{3\pi}{4\sqrt{2}}\sqrt{\delta_{\ell}}\,\,
\frac{{\cal H}_{P}-2\,{\cal H}_{SL}}{{\cal H}_{\rm tot}}.
\label{eq:Pol_lept}
\ena
The corresponding $\theta$ dependent components can be obtained from
Appendix A.
The longitudinal polarization $P^\ell_z(q^2)$ can be seen to have the rate 
structure 
$(\Gamma(hf)-\Gamma(nf))/(\Gamma(hf)+\Gamma(nf))$ in agreement with 
Eq.~(\ref{eq:Pol_lept}).
Note that the scalar-longitudinal piece ${\cal H}_{SL}$ in~(\ref{eq:Pol_lept}) 
contains an extra minus sign resulting from the factor $(-1)^{J+J'}$.

When calculating the $q^{2}$ averages of the components of
$\vec P^{h}$and $\vec P^{\ell}$, one has to reinstate the common 
$q^{2}$--dependent factor
$(q^2-m_\ell^2)^2 |\mathbf{p_2}|/q^2$ in the numerator and denominator of
the right-hand sides of Eqs.~(\ref{eq:Pol_had}) and (\ref{eq:Pol_lept}).

\section{Fourfold angular decay distribution}

In the previous section, we have calculated the polarization of the
charmed baryon $\Lambda_{c}^{+}$ in the three-body decay
$\Lambda_b^0 \to \Lambda_c^+ +\ell^- \, \bar \nu_\ell$. The polarization of
the $\Lambda_{c}^{+}$ can be probed by analyzing the angular decay 
distribution of the subsequent decay of the $\Lambda_{c}^{+}$. As an 
exemplary case we shall consider the decay mode 
$\Lambda_{c}^{+}\to \Lambda^0+\pi^+$ as polarization analyzer. We shall 
discuss some other
important decay modes of the $\Lambda_{c}^{+}$ at the end of this section.

One  can exploit the cascade nature of the decay
$\Lambda_b^0 \to \Lambda_c^+ (\to \Lambda^0+\pi^+) 
+ W^-_{\rm off-shell}(\to \ell^- + \bar \nu_\ell)$
by writing down a joint angular decay distribution involving the polar angles 
$\theta,\, \theta_{B}$ and the azimuthal angles $\chi$ defined by the
decay products in their respective CM (center of mass) systems as shown in 
Fig.~\ref{fig:angles}. The angular decay distribution involves
the helicity amplitudes 
$H_{\lambda_{\Lambda_{c}}\lambda_{W}}$ for the decay 
$\Lambda_b \to \Lambda_c + W$ and $h^{B}_{\lambda_{\Lambda}0}$ 
for the decay $\Lambda^+_c \to \Lambda^0 + \pi^+$. The joint angular decay
distribution for the decay of an unpolarized $\Lambda_b$ reads
\bea
W(\theta,\theta_B,\chi)  & = &
\sum_{\lambda_W,\lambda'_W,\lambda_2,\lambda'_2,\lambda_3}
 (-1)^{J+J'} 
d^{1/2}_{\lambda_2\lambda_3}(\theta_B)
d^{1/2}_{\lambda'_2\lambda_3}(\theta_B)
h^B_{\lambda_3 0}h^{B\,\dagger}_{\lambda_3 0}
\nonumber \\ 
&\times&
\delta_{\lambda_W - \lambda_2,\lambda'_W-\lambda'_2}
H_{\lambda_2\lambda_W}\,
H^{\dagger}_{\lambda'_2\lambda'_W}\,
L_{\lambda_{W}\,\lambda'_{W}}(\theta,\chi)
\label{eq:fourfold1}
\ena
where the $\theta$-- and $\chi$--dependent leptonic helicity tensor 
$L_{\lambda_{W}\,\lambda'_{W}}(\theta,\chi)$ is given in Appendix A.

For computer processing, we have taken a slight variant of~(\ref{eq:fourfold1})
where we drop the ``$\lambda_{W}=t$'' convention and make up for this by 
introducing an 
explicit $(J,J')$ dependence for the hadronic and leptonic helicity 
amplitudes. Also, we express the leptonic helicity tensor 
$L_{\lambda_{W}\,\lambda'_{W}}(\theta,\chi)$ in terms of rotated
products of leptonic helicity similar to~(\ref{Lhat}) but now including also 
the azimuthal rotation. One has
\be
L_{\lambda_{W}\,\lambda'_{W}}(\theta,\chi)=
d^J_{\lambda_W,\lambda_{\ell}-1/2}(\theta)\,
d^{J'}_{\lambda'_W,\lambda_{\ell}-1/2}(\theta)\,
e^{i(\lambda_W-\lambda'_W)\chi}\,
h_{\lambda_\ell,\lambda_{\bar\nu_\ell=1/2}}\,(J)\,
h_{\lambda_\ell,\lambda_{\bar\nu_\ell=1/2}}\,(J') .
\label{rot2}
\en
Our master formula for the general angular decay distribution then reads
\bea
W(\theta,\theta_B,\chi)  & = &
\sum\limits_{\rm set}
 (-1)^{J+J'} 
d^{1/2}_{\lambda_2\lambda_3}(\theta_B)\,
d^{1/2}_{\lambda'_2\lambda_3}(\theta_B)\,
h^B_{\lambda_3 0}\,h^{B\,\dagger}_{\lambda_3 0}
\nonumber \\ 
&\times&
(\delta_{\lambda_W-\lambda_2,\,+1/2}
+\delta_{\lambda_W-\lambda_2,\,-1/2})
\delta_{\lambda_W - \lambda_2,\lambda'_W-\lambda'_2}
H_{\lambda_2\lambda_W}\,(J)\,
H^{\dagger}_{\lambda'_2\lambda'_W}\,(J')
\nonumber \\
&\times&
d^J_{\lambda_W,\lambda_{\ell}-1/2}(\theta)
d^{J'}_{\lambda'_W,\lambda_{\ell}-1/2}(\theta)
e^{i(\lambda_W-\lambda'_W)\chi}
h_{\lambda_\ell,\lambda_{\bar\nu_\ell=1/2}}\,(J)
h_{\lambda_\ell,\lambda_{\bar\nu_\ell=1/2}}\,(J') .
\label{eq:fourfold2}
\ena
The summation is performed over the following
set of the indices: 
\be
\text{set}=\Big\{
\lambda_{\ell},
\lambda_2,\lambda'_2,
\lambda_3 = \pm\frac12,\quad
\lambda_W,\lambda'_W =\pm 1, 0,\quad
J,J'=0,1
\Big\},
\label{set}
\en
where, in the present application, $B_{1}=\Lambda_{b},\,B_{2}=\Lambda_{c}$ 
and $B_{3}=\Lambda$. Note that the set~(\ref{set}) does not contain
a summation over $\lambda_{W}=t$ because this is now replaced by the 
sum over $J,J'$. As before, the Kronecker symbol 
$\delta_{\lambda_W - \lambda_2,\lambda'_W - \lambda'_2}$ in
Eq.~(\ref{eq:fourfold1}) expresses the fact that we are considering 
the decay of an unpolarized $\Lambda_b$. We have also introduced a
second Kronecker symbol in Eq.~(\ref{eq:fourfold1}), which expresses the
fact that the parent baryon is a spin-$1/2$ particle, implying 
$|\lambda_W - \lambda_2|=|\lambda'_W-\lambda'_{2}|=1/2$. 
The latter condition is expressed through the Kronecker symbol 
$(\delta_{\lambda_W - \lambda_2,\,+1/2}
 +\delta_{\lambda_W - \lambda_2,\,-1/2})$. 
The $d^j_{mm'}$ with $(j=0,1/2,1)$ are 
Wigner's $d$--functions, where $d^{0}_{00}=1$. We mention that it is not 
difficult to generalize Eq.~(\ref{eq:fourfold1}) to the case of a decaying 
polarized $\Lambda_b$, as has been done in corresponding baryonic 
cascade decay calculations 
in Refs.~\cite{Gutsche:2013oea,Bialas:1992ny,Kadeer:2005aq,Fischer:2003vu,
Fischer:1998gsa,Fischer:2001gp}. 

After factoring out the $\Lambda_{c}^{+}\to \Lambda^0 +\pi^{+}$ rate term
$(|h^B_{+\frac12\,0}|^2 \,+\,  |h^B_{-\frac12\,0}|^2)$, one obtains
\bea
W(\theta,\theta_B,\chi)&=&\frac{8 q^2 v}{3}\Big(|h^B_{+\frac12\,0}|^2 \,
+\,  |h^B_{-\frac12\,0}|^2\Big)\quad \bigg\{ 
   \frac{3}{8}(1+\cos^{2}\theta)\,{\cal H}_{U}
 \mp \frac 34 \cos\theta \,{\cal H}_{P}+\frac{3}{4}\sin^2\theta \,{\cal H}_{L}
\nn 
&+& \delta_{\ell}\,\Big(\frac 34 \sin^{2}\theta \,{\cal H}_{U}+
\frac 32 \cos^{2}\theta\,{\cal H}_{L}-3\cos\theta\,{\cal H}_{SL}
+\frac{3}{2}\,{\cal H}_{S}
\Big)
 \nn
&+&\alpha_{B} \cos\theta_{B}
\bigg[\frac{3}{8}(1+\cos^{2}\,\theta){\cal H}_{P} \,
 \mp \frac 34 \cos\theta\, {\cal H}_{U} +\frac{3}{4}\sin^2\theta\,
{\cal H}_{L_{P}}
 \nn
&+&\delta_{\ell}\,\Big(\frac 32 \cos^{2}\theta \,{\cal H}_{L_P}
+\frac 34 \sin^{2}\theta\,{\cal H}_{P}+\frac 32 {\cal H}_{S_{P}}
-3\cos\theta\,{\cal H}_{SL_{P}}
\Big)
\bigg] 
 \nn
&+&\alpha_B\sin\theta_B \cos\chi
 \bigg[\mp \frac{3}{2\sqrt{2}} \sin\theta
 \,{\cal H}_{LT} + \frac{3}{4\sqrt{2}} \sin2\theta  \,{\cal H}_{LT_{P}}
 \nn
&+&\delta_{\ell}\,\Big(-\frac{3}{2\sqrt{2}}\sin2\theta\,{\cal H}_{LT_{P}}
+\frac{3}{\sqrt{2}}\sin\theta\,{\cal H}_{ST_{P}}
\Big)
\Big]
\bigg\}\,, 
\label{eq:fourfold3}
\ena
where the polarization asymmetry $\alpha_B $ of the decay 
$\Lambda_{c}^+ \to \Lambda^0 + \pi^{+}$
is defined by
\be
\alpha_B = \frac{ |h^B_{+\frac12\,0}|^2 \,-\,  |h^B_{-\frac12\,0}|^2}
                { |h^B_{+\frac12\,0}|^2 \,+\,  |h^B_{-\frac12\,0}|^2}\,.
\label{eq:alphaB}
\en

In order to be able to also discuss the decays 
$\Lambda^+_c \to \Lambda^0 + \ell^+ + \nu_\ell$ and 
$\bar \Lambda^0_b \to \bar\Lambda^-_c + \ell^+ + \nu_\ell$, we have
included the necessary sign changes in Eq.~(\ref{eq:fourfold3}). 
The upper sign refers to the 
lepton configuration $(\ell^-,\bar{\nu}_\ell)$,  
and the lower sign refers to the $(\ell^+,\nu_\ell)$ configuration. 
The changes are effected by replacing 
$d^J_{\lambda_W,\lambda_{\ell}-1/2}(\theta) \to 
d^J_{\lambda_W,\lambda_{\ell}+1/2}(\theta)$
and
$h_{\lambda_\ell,\lambda_{\bar\nu_\ell=1/2}}
\to h_{\lambda_\ell,\lambda_{\nu_\ell=-1/2}}$ in Eq.~(\ref{eq:fourfold2}).
The sign changes in Eq.(\ref{eq:fourfold3}) can be seen to result from 
the parity-violating contribution of the lepton tensor, as discussed in 
Appendix A. 

Let us briefly pause to discuss the cascade decay 
$\bar \Lambda^0_b \to \bar\Lambda^-_c (\to \bar \Lambda^0 +\pi^{-}) 
+ \ell^+ + \nu_\ell$ and how the
angular decay distribution of the charge conjugated mode is related to that of 
$\Lambda^0_b \to \Lambda^+_c (\to \Lambda^0 +\pi^{+})+ \ell^- +\bar \nu_\ell$. 
We exploit the $CP$ properties of the relevant helicity structure functions. 
One has
\be
{\cal H}_{X}\,^{\mathrm{pc}/\mathrm{pv}}\,(\bar \Lambda_{b})
=\pm {\cal H}_{X}\,^{\mathrm{pc}/\mathrm{pv}}\,(\Lambda_{b})
\qquad {\rm and}\qquad \alpha_{\bar B}= - \alpha_{B}\,.
\en
This shows that the angular decay distribution of the charge conjugated
mode is identical to the original mode as long as one replaces the particles
in the decay chain 
$\Lambda^0_b \to \Lambda^+_c (\to \Lambda +\pi^{+})+ \ell^- +\bar \nu_\ell$ 
with their charge conjugates.

Equation~(\ref{eq:fourfold3}) contains 11 distinct helicity structure 
functions which can be 
measured by an angular analysis of the cascade decay 
$\Lambda_b^0 \to \Lambda_c^+ (\to \Lambda^0+\pi^+) 
+ W^-_{\rm off-shell}(\to \ell^- + \bar \nu_\ell)$. Some of the 
helicity structure functions multiply the same angular factors. To separate
them, one also has to take into account the dependence on the factor
$\delta_\ell(q^{2})$.

In Eq.~(\ref{eq:fourfold1}) we have summed over the helicity labels of 
the lepton; i.e., we have taken the trace of the respective spin density 
matrix.
By leaving the respective helicity labels unsummed, one can then obtain
the $(\theta,\theta_{B},\chi)$--dependent polarization of the lepton
including the result Eq.~(\ref{eq:Pol_lept}) after integrating 
Eq.~(\ref{eq:fourfold2}) over $\cos\theta,\,\cos\theta_{B}$ and $\chi$.

Using the narrow width approximation for the intermediate baryon state 
$\Lambda^{+}_{c}$ one finally obtains
\bea
\frac{d\Gamma(\Lambda_b^0\to\Lambda^{+}_c(\to 
\Lambda\pi^{+})+\ell\bar\nu_\ell)}
     {dq^2 d\cos\theta d\chi d\cos_B} &=&
\frac12 \frac{1}{2\pi}
\mathrm{Br}(\Lambda^{+}_c \to \Lambda+\pi^{+})\Gamma_{0}
\frac{(q^2-m_l^2)^2|\mathbf{p_2}|}{M_1^7q^2}  
\nonumber \\[2ex]
&\times&W(\theta) \,
\Big(1 + P^{h}_z(\theta)\, \alpha_B\cos\theta_B  
+ P^{h}_x(\theta)\, \alpha_B\sin\theta_B\cos\chi \Big)\,, 
\label{eq:fourfold4}
\ena
where we have factored out the unpolarized
rate expression $W(\theta)=3/8\,(2q^{2}v^{2})^{-1}H^{\mu\nu}L_{\mu\nu}(\theta)$
defined in Eq.~(\ref{two1}). $P^{h}_{z}(\theta)$ and  
$P^{h}_{x}(\theta)$ are the
$\theta$--dependent hadron-side polarization components which read 
\bea
P^{h}_{z}(\theta)&=&
\frac{1}{W(\theta)}\bigg[\frac{3}{8}(1+\cos^{2}\,\theta){\cal H}_{P} \,
\mp\frac 34 \cos\theta\, {\cal H}_{U} 
+\frac{3}{4}\sin^2\theta\,{\cal H}_{L_{P}}
\nonumber \\[-2ex]
&&\qquad\qquad\qquad+\delta_{\ell}\Big(\frac 32 \cos^{2}\theta \,
{\cal H}_{L_{P}}
+\,\frac 34 \sin^{2}\theta\,{\cal H}_{P}+\frac 32 {\cal H}_{S_{P}}
-3\cos\theta\,{\cal H}_{SL_{P}}
\Big)\bigg]\,, 
\nonumber\\
P^{h}_{x}(\theta)&=&\frac{1}{W(\theta)}\bigg[\mp \frac{3}{2\sqrt{2}} 
\sin\theta \,{\cal H}_{LT} +\frac{3}{4\sqrt{2}} \sin2\theta  
\,{\cal H}_{LT_{P}}
+\delta_{\ell}\Big(-\frac{3}{2\sqrt{2}}\sin2\theta\,{\cal H}_{LT_{P}}
+\frac{3}{\sqrt{2}}\sin\theta\,{\cal H}_{ST_{P}}
\Big)\bigg]\,. 
\label{eq:Pol_had1}
\ena
When integrating the polarization components $P^{h}_{z}(\theta)$ and  
$P^{h}_{x}(\theta)$ in  Eq.~(\ref{eq:Pol_had1}) (numerator and denominator
separately), one recovers the corresponding expressions 
in Eq.~(\ref{eq:Pol_had}).

Since our model amplitudes are real, and since we are not considering
$CP$--violating effects inasmuch as $V_{cb}$ is real, we have omitted a 
possible contribution
proportional to $P_y(\theta)\, \alpha_B\sin\theta_B\sin\chi$ in the second 
line of the distribution~(\ref{eq:fourfold4}). The perpendicular
polarization component $P_y(\theta)$ is contributed to by the 
absorptive counterparts of the dispersive structure functions ${\cal H}_{LT},
\,{\cal H}_{LT_{P}}$ and ${\cal H}_{ST_{P}}$. They are
${\rm Im}(  H_{+\frac12 +1}  H_{-\frac12\, 0}^\dagger 
                         + H_{+\frac12\, 0} H_{-\frac12 -1}^\dagger )$, 
${\rm Im}(  H_{+\frac12 +1}  H_{-\frac12\, 0}^\dagger 
                           - H_{+\frac12\, 0} H_{-\frac12 -1}^\dagger )$
and ${\rm Im}(  H_{+\frac12 +1} H_{-\frac12\,t}^\dagger 
                          - H_{+\frac12\,t} H_{-\frac12 -1}^\dagger )$, 
including also possible $CP$--odd phases.
The three absorptive observables correspond to $T$--odd observables. The 
angular factors projecting onto them can be seen to correspond to 
$T$--odd momentum products (see Ref.~\cite{Kadeer:2005aq}). We emphasize that
there are no absorptive parts originating from our two-loop calculation 
since we have incorporated confinement from the beginning; i.e., the 
perpendicular polarization component $P_y(\theta)$ is zero in our model. 
 
The last term in the round bracket of Eq.~(\ref{eq:fourfold4}) proportional 
to $P^{h}_{x}(\theta)$ describes the
azimuthal dependence of the decay rate.
After $\cos\theta_{B}$-- and $\cos\theta$--integration 
Eq.~(\ref{eq:fourfold4}) allows one 
to define an azimuthal asymmetry parameter
\be
\frac{d\Gamma}{dq^{2}d\chi}\propto (1+\alpha_{B}\,\gamma(q^{2}) \cos\chi)
\label{azipar}
\en
where
\be
\gamma(q^{2})= -\frac{3\pi^{2}}{16\sqrt{2}}
\frac{({\cal H}_{LT}-2\delta_{\ell}{\cal H}_{ST_{P}})}{{\cal H}_{tot}}
=\frac{\pi}{4}P^{h}_{z}(q^{2}) .
\label{azipar1}
\en

The advantage of the $\Lambda_{c}^{+}\to \Lambda^0+\pi^+$ mode discussed
in this section is that its analyzing power is close to maximal with 
$\alpha_{B}=-0.91\pm0.15$ \cite{Agashe:2014kda}. A disadvantage of this mode 
is that the $\Lambda^0$ is rather long lived, with a lifetime of 
$\tau_{\Lambda}=2.63\cdot 10^{-10}s$ \cite{Agashe:2014kda}, meaning
that at the LHC many of the final state $\Lambda$'s decay outside of the 
e.g. LHCb detector such that the decays $\Lambda^0 \to p\pi^{-},n\pi^{0} $ 
cannot always be reconstructed. This will become even worse in run 2 of the 
LHC when the energy, and thus the average energy of the final-state 
$\Lambda$'s, will increase. The optimal place to look for the decay chain 
$\Lambda_{b}\to \Lambda_{c}\to \Lambda$ would be for $\Lambda_{b}$'s  
produced at the GigaZ@ILC option.  

Other decay channels of the $\Lambda_{c}$ with comparable branching ratios
larger than $1\,\%$ are  
$\Lambda_{c}\to \Lambda \ell^{+}\nu 
\,(-0.86 \pm 0.04\,$\cite{Agashe:2014kda}),
$\,p\bar K^{0} (-1\,$\cite{Korner:1992wi}),
$\,\Sigma^{+}\pi^{0}(-0,45\pm0.31 \pm0.06\,$\cite{Agashe:2014kda};
\,0.71\,\cite{Korner:1992wi}),
$\,\Sigma^{0}\pi^{+}(+0.70\,$\cite{Korner:1992wi}), 
$\,p\bar K^{\ast\,0}(+0,69\,$\cite{Konig:1993wz}),
$\,pK^{-}\pi^{+}$, $ \Lambda\rho^{+},\,\Sigma^{+}\rho^{0}$, 
$\Sigma^{0}\rho^{+}$.
We have added experimental values (with errors) and theoretical values 
(without errors) for the respective polarization asymmetry parameters when available.
The analyzing power of some of these other modes is not as large as that of the
$\Lambda_{c}^{+}\to \Lambda^0+\pi^+$ mode but may still lead to reliable
measurements of the longitudinal polarization of the~$\Lambda_{c}^{+}$.

\section{The transition form factors in the covariant confined quark model}

We shall use the covariant confined quark model previously developed by us
to describe the dynamics of the current--induced $\Lambda_b = (b[ud])$ to
$\Lambda_c = (c[ud])$ transition 
(see Refs.~\cite{Gutsche:2013oea,Gutsche:2013pp,Gutsche:2012ze}).
The starting point of the model is an interaction Lagrangian
which describes the coupling of the $\Lambda_Q$-baryon to
the relevant interpolating three-quark current. One has
\bea
{\cal L}^{\,\Lambda_Q}_{\rm int}(x) 
&=&g_{\Lambda_Q} \,\bar\Lambda_Q(x)\cdot J_{\Lambda_Q}(x) 
 + g_{\Lambda_Q} \,\bar J_{\Lambda_Q}(x)\cdot \Lambda_Q(x)\,,  
\label{eq:Lagr}\\[2ex]
J_{\Lambda_Q}(x) &=& \int\!\! dx_1 \!\! \int\!\! dx_2 \!\! \int\!\! dx_3 \, 
F_{\Lambda_Q}(x;x_1,x_2,x_3) \, J^{(\Lambda_Q)}_{3q}(x_1,x_2,x_3)\,,
\nn
J^{(\Lambda_Q)}_{3q}(x_1,x_2,x_3) &=& 
 \epsilon^{a_1a_2a_3} \, Q^{a_1}(x_1)\, u^{a_2}(x_2) 
\,C \, \gamma^5 \, d^{a_3}(x_3)\,,
\nn[2ex]
\bar J_{\Lambda_Q}(x) &=& 
\int\!\! dx_1 \!\! \int\!\! dx_2 \!\! \int\!\! dx_3 \, 
F_{\Lambda_Q}(x;x_1,x_2,x_3) \, \bar J^{(\Lambda_Q)}_{3q}(x_1,x_2,x_3)\,,
\nn
\bar J^{(\Lambda_Q)}_{3q}(x_1,x_2,x_3) &=& 
\epsilon^{a_1a_2a_3} \, \bar d^{a_3}(x_3)\, \gamma^5 \,C\, \bar u^{a_2}(x_2)  
\cdot \bar Q^{a_1}(x_1)\,. 
\nonumber
\ena
The vertex function $F_{\Lambda_Q}$ is chosen to be of the form
\be
F_{\Lambda_Q}(x;x_1,x_2,x_3) \, = \, 
\delta^{(4)}(x - \sum\limits_{i=1}^3 w_i x_i) \;  
\Phi_\Lambda\biggl(\sum_{i<j}( x_i - x_j )^2 \biggr) 
\label{eq:vertex}
\en 
where $\Phi_{\Lambda_Q}$ is a correlation function involving the three 
constituent quarks with coordinates $x_1$, $x_2$, $x_3$ and with 
masses $m_1$, $m_2$, $m_3$. The variable $w_i$ is defined by 
$w_i=m_i/(m_1+m_2+m_3)$ such that $\sum_{i=1}^3 w_i=1.$ 
The form factors describing the $\Lambda_Q\to\Lambda_{Q'}$ transition
via the local weak quark current are calculated in terms of a two-loop 
Feynman diagram.
Due to the confinement mechanism of the model, the Feynman
diagrams do not contain branch points corresponding to on-shell quark 
production.

The values of the constituent quark masses $m_q$ are taken from a new 
fit which improves the description of new data on the exclusive  
$B$-meson and the $\Lambda_c$ decays. In the fit, the infrared cutoff 
parameter $\lambda$ of the model has been kept fixed. One has 
\be
\def\arraystretch{2}
\begin{array}{ccccccc}
     m_u        &      m_s        &      m_c       &     m_b & \lambda  &   
\\\hline
 \ \ 0.241\ \   &  \ \ 0.428\ \   &  \ \ 1.67\ \   &  \ \ 5.04\ \   & 
\ \ 0.181\ \   & \ {\rm GeV} 
\end{array}
\label{eq: fitmas}
\en
The values of the size parameters are taken from our previous papers
\cite{Gutsche:2013pp,Gutsche:2013oea}:
\be
\def\arraystretch{2}
\begin{array}{cccc}
 \Lambda_{\Lambda_s}  &   \Lambda_{\Lambda_c} &  \Lambda_{\Lambda_b}  &   
\\\hline
   \ \ 0.492 \ \   &  \ \ 0.867\ \       &  \ \ 0.571 \ \     & \ {\rm GeV} 
\end{array}
\label{eq: size}
\en

\begin{figure}
\begin{center}
\hspace*{-0.5cm}
\begin{tabular}{lr}
\includegraphics[width=0.45\textwidth]{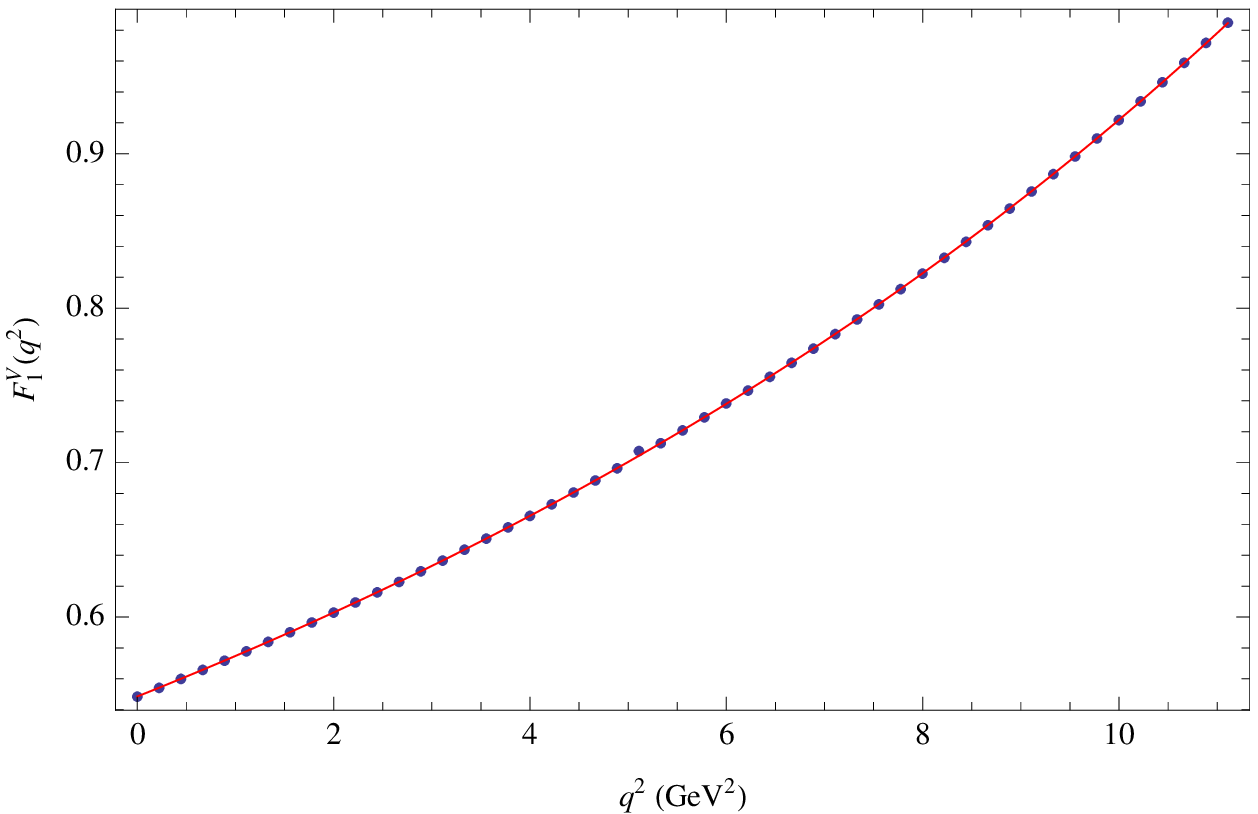}   & \hspace*{.5cm}
\includegraphics[width=0.45\textwidth]{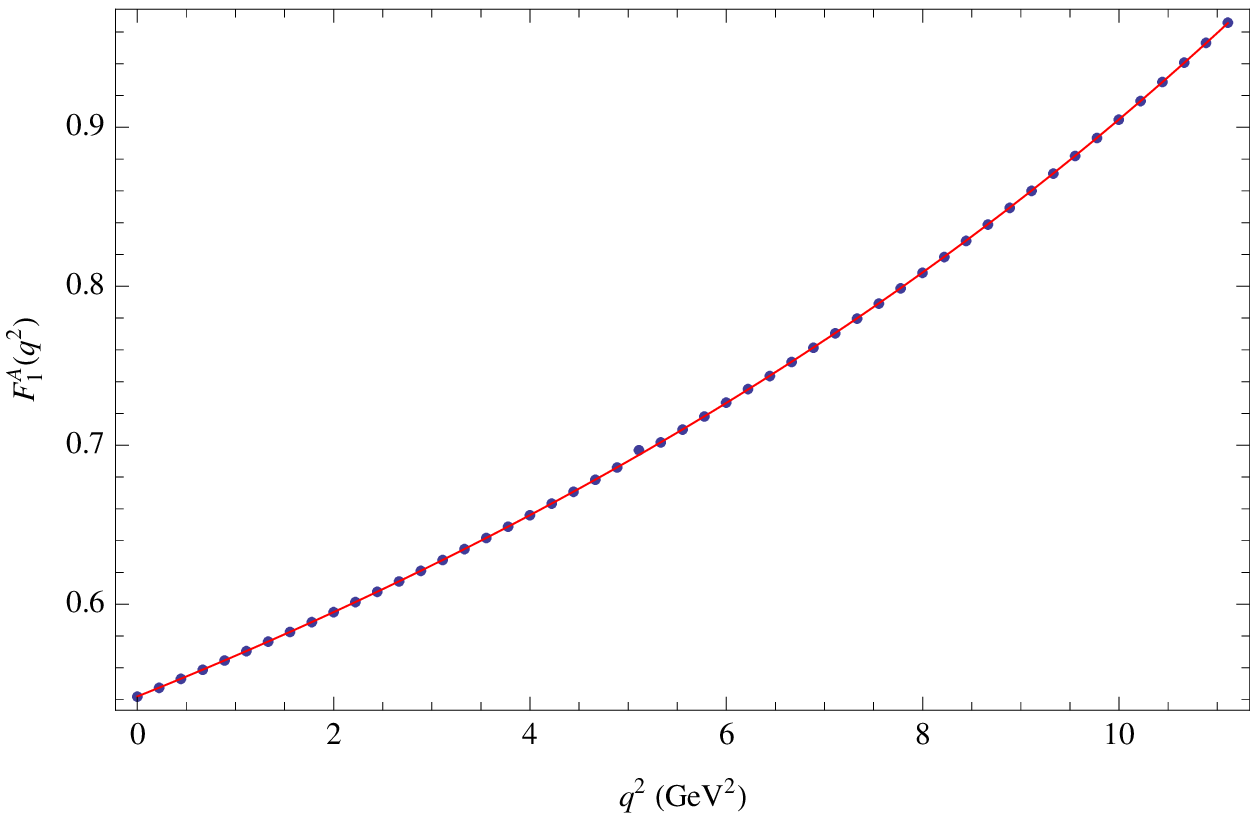}      \\[2ex]

\includegraphics[width=0.45\textwidth]{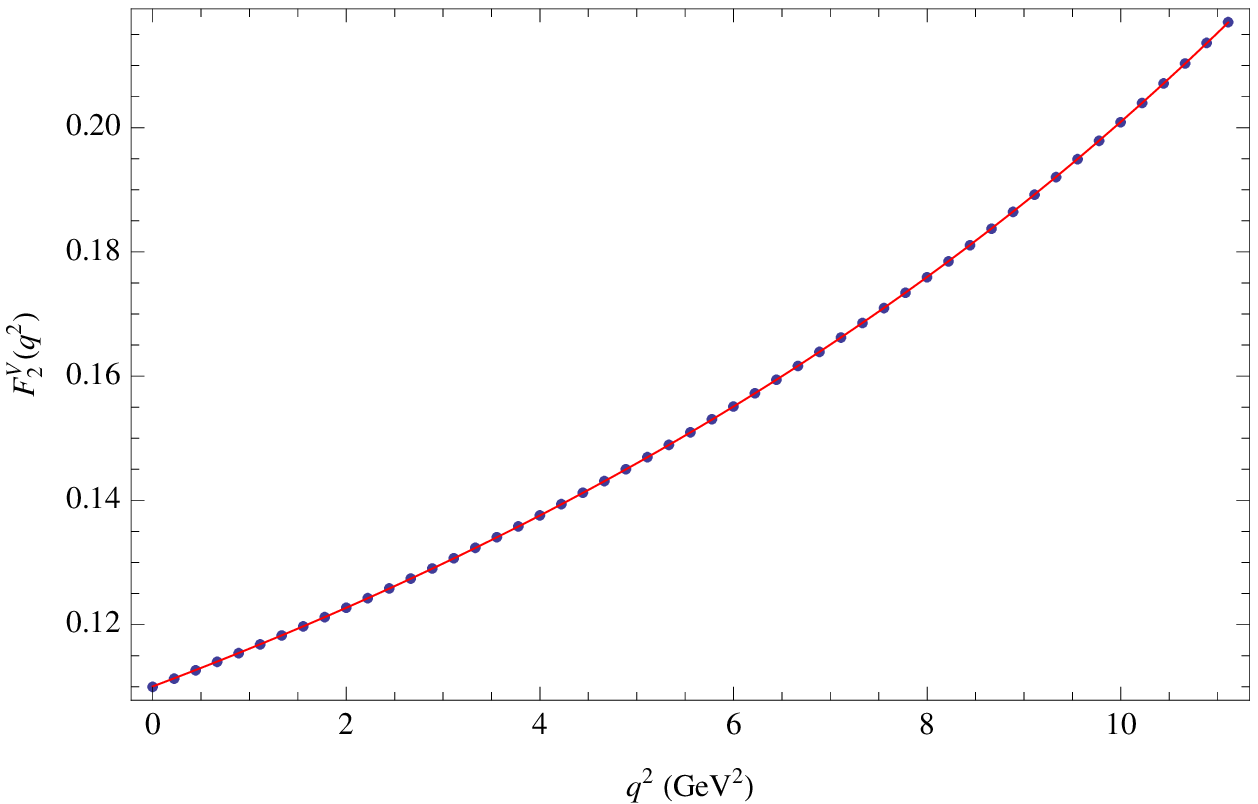}  & \hspace*{.5cm}
\includegraphics[width=0.45\textwidth]{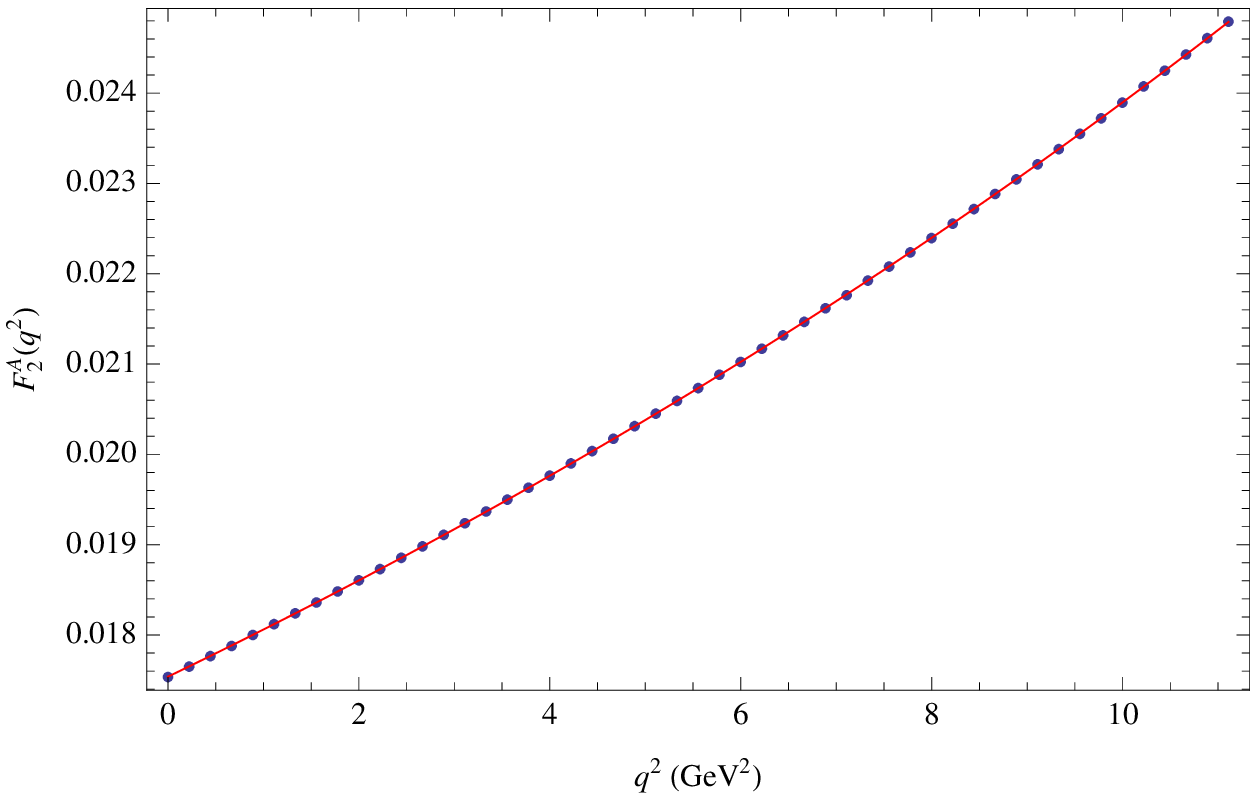}    \\[2ex]

\includegraphics[width=0.45\textwidth]{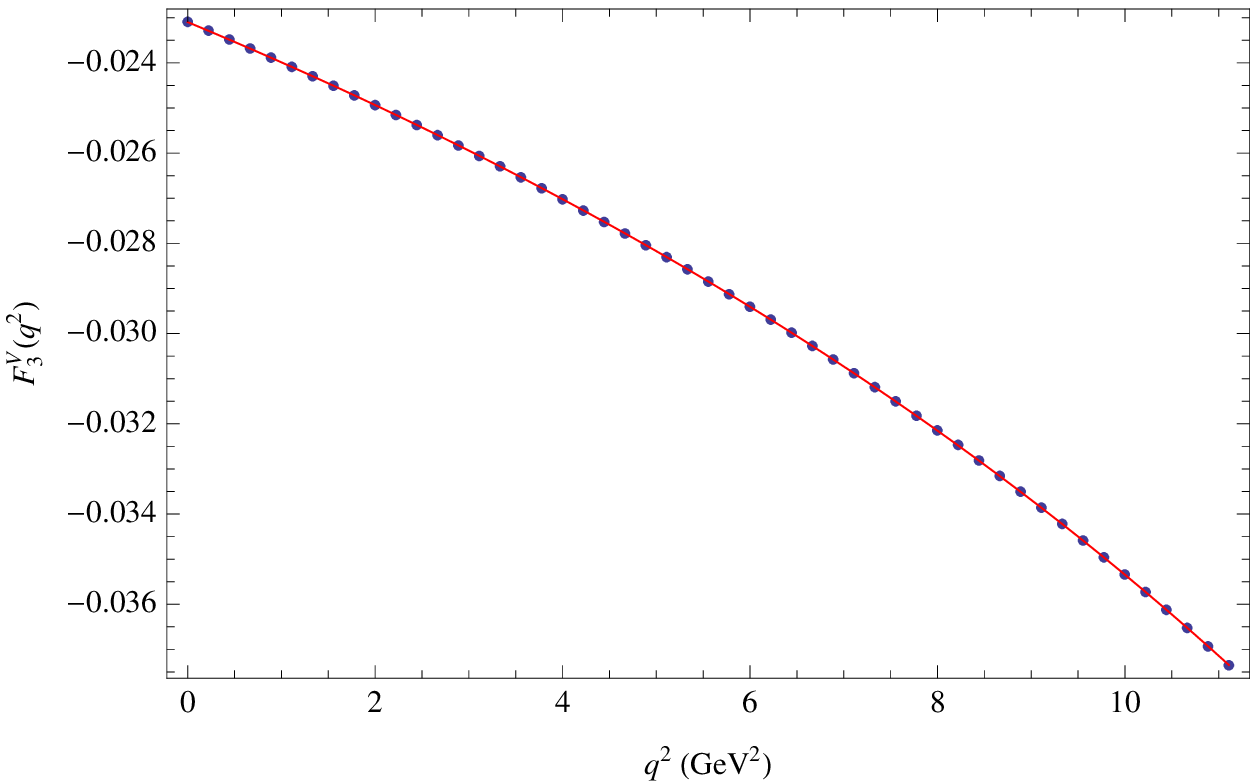}  & \hspace*{.5cm}
\includegraphics[width=0.45\textwidth]{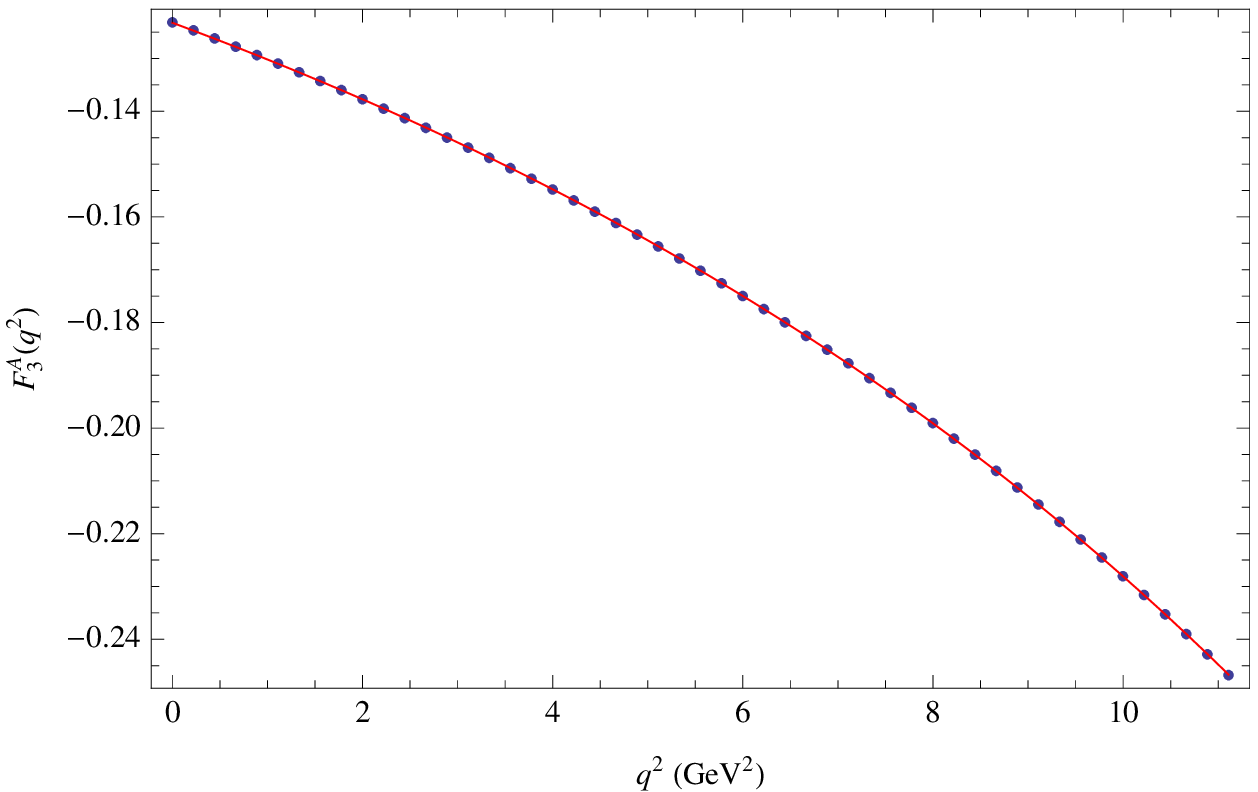}
\end{tabular}
\caption{\label{fig:ff_bc}
Form factors defining the transition $\Lambda_b\to\Lambda_c$:
approximated results (solid line), exact result(dotted line).
}
\end{center}
\end{figure}

The results of our numerical two-loop calculation are well represented
by a double--pole parametrization
\be\label{eq:DPP}
F(q^2)=\frac{F(0)}{1 - a s + b s^2}\,, \quad s=\frac{q^2}{M_1^2} 
\en 
with high accuracy: the relative error is less than 1\%. 
In Fig.~\ref{fig:ff_bc} we display a comparison of the 
exact results for the relativistic form factors $F_i^V$ and $F_i^A$ 
$(i=1,2,3)$ with the double-pole parametrization. 
For the $\Lambda_b \to \Lambda_c$ transition the parameters of the 
approximated form of the form factors are given by
\be
\begin{array}{ccccccc}
 &\quad F_1^V \quad &\quad  F_2^V \quad & \quad F_3^V \quad 
 &\quad F_1^A \quad & F_2^A \quad & F_3^A \quad
\\[1.5ex]
\hline\\[-2ex]
F(0) &  0.549  & 0.110 & -0.023 & 0.542 & 0.018 & -0.123
\\[1ex] 
a    &  1.459  & 1.680 &  1.181 & 1.443 & 0.921 &  1.714
\\[1ex] 
b    &  0.571  & 0.794 &  0.276 & 0.559 & 0.255 &  0.828
\\[1.1ex] 
\hline
\end{array}
\label{eq:ff_param}
\en
The dominant form factors are $F_1^V(q^{2})$ and $F_1^A(q^{2})$ with very 
similar $q^{2}$ dependencies. Inasmuch as one can neglect the other form
factors, the results of our model calculation are very close to the 
leading-order HQET result
$F_{1}^{V}(q^{2})=F_{1}^{A}(q^{2})=F(q^{2}),\,F_{2,3}^{V/A}=0$ discussed
in Sec.~II. In fact, our numerical results to be discussed in Sec.~VII
are quite close to what would be expected in leading-order HQET.   

Let us take a closer look at the $q^2$ dependence of the
form factors $F_1^V(q^{2})\approx F_1^A(q^{2}) $. Their $q^2$ dependence is 
very close to a dipole behavior,  
since one has $\sqrt{b} \sim a/2$ in both cases with a dipole 
mass $m_{\rm dipole} = M_1/\sqrt{a/2} \sim 6.6$ GeV. The dipole mass is
quite close to the expected $(b\bar c)$ mass scale of $6.28$ GeV set by the 
$B_{c}$-meson mass~\cite{Agashe:2014kda}. Next, we look at the 
near zero recoil behavior of $F_1^V(q^{2})$ and $F_1^A(q^{2})$. In order to
investigate the near recoil zero behavior of the form factors, we switch to 
the variable $w = (M_1^2 + M_2^2 -q^2)/(2M_1M_2)$ such that $w=1$ at the 
zero recoil point $q^2=q^2_{\rm max}=(M_1-M_2)^2$.
The Taylor expansion of any given function $f(w)$ around the zero recoil 
point $w=1$
reads
\bea
f(w)&=&\sum\limits_{n=0}^\infty \frac{f^{(n)}}{n!}(w-1)^n
\nn
    &=&
f(1)\,\Big\{1 + \frac{f'(1)}{f(1)}(w-1)+\frac{f^{''}(1)}{2 f(1)}(w-1)^2 + \ldots
      \Big\}
\nn
&\equiv&
f(1)\,\Big\{1 - \rho^2 (w-1)+c\,(w-1)^2 + \ldots
      \Big\}
\label{eq:w-expansion}
\ena
where $\rho^2$ is called the slope parameter and $c$ 
is the convexity parameter.
It is not difficult to express the parameters $f(w=1)$, $\rho^2$ and $c$ via
the original parameters $F(0)$, $a$ and $b$ used in the double--pole
parametrization Eq.~(\ref{eq:DPP}). One finds
\bea
f(w=1) &=& \frac{F(0)}{1-a\,s_{\rm max} + b\,s_{\rm max}^2}\,,
\nn[1.5ex]
\rho^2 &=& \frac{2\,r\, (a-2\, b\, s_{\rm max})} 
                {1-a\,s_{\rm max} + b\,s_{\rm max}^2}\,,
\qquad
c =  \frac{4\,r^2\,
    \left[a^2 - b -3\, a\, b\, s_{\rm max} + 3\, b^2\, s_{\rm max}^2\right]}
           {\left[1-a\,s_{\rm max} + b\,s_{\rm max}^2\right]^2}
\label{eq:w-param}
\ena
where $r=M_2/M_1$ and $s_{\rm max}=(1-r)^2$. The numerical results 
for $F_1^V(q^{2})$ and $F_1^A(q^{2})$ can be calculated to be 

\be 
\begin{array}{ccccccc}
                   & \quad   f(w=1)\quad & \quad \rho^2 \quad & \quad c \quad
\\[1.5ex]
\hline \\[-2ex]
\quad F_1^V \quad  &       0.985       &        1.543       &       1.704
\\[1.2ex]
\quad F_1^A \quad  &       0.966       &        1.521       &       1.654
\\[1.1ex] 
\hline
\end{array}
\en
First, one notes that the zero recoil normalization is very close to 1
for both $F_1^V$ and $F_1^A$, which would be the normalization predicted
by leading order HQET. The values for the two slope parameters are compatible
with the only experimental result published by the DELPHI Collaboration:  
$\rho^{2}=2.03\pm0,46({\rm stat})^{+0.72}_{-1.00}
({\rm syst})$~\cite{Abdallah:2003gn}. 
There are no experimental results on the convexity
parameter yet. There are a number of theoretical model calculations for
the slope parameter of the $\Lambda_{b} \to \Lambda_{c}$ transitions, many of 
which scatter around $\rho^{2}\approx 1.5$~\cite{Konig:1993ze,Ivanov:1996fj,%
Korner:2000zn,Huang:2005mea,Pervin:2005ve,%
Ebert:2006rp,Ke:2007tg,Grozin:1992mk,Woloshyn:2014hka}. 
A value around 1.5 is expected from the spectator quark model relation 
$\rho^{2}_{B}=2\rho^{2}_{M}-1/2$~\cite{Hussain:1990uu} using 
a mesonic slope parameter of $\rho^{2}_{M}\approx 1$ as reported
by the Heavy Flavor Averaging Group~\cite{Amhis:2014hma}.

\section{Numerical results}

We shall present numerical results for the two cases $\ell^{-}=e^{-}$ 
and $\ell^{-}=\tau^{-}$. The results for the $\mu^{-}$ mode are almost
identical to those of the $e^{-}$ mode and will not be listed separately. 
The dynamical input in terms of the covariant confined quark model has 
been specified in Sec.~VI. We use the mass values 
$M_{\Lambda_{b}}=5.6195$ GeV, $M_{\Lambda_{c}}=2.2864$ GeV and
$m_{\tau}=1.7768$ GeV~\cite{Agashe:2014kda}. 

\begin{figure}[ht]
\begin{center}
\vspace*{.25cm}
\epsfig{figure=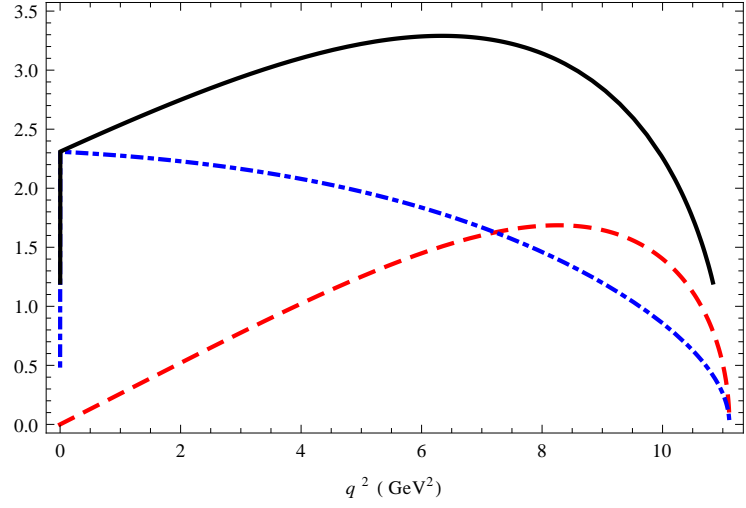,scale=.8}
\caption{The  $q^{2}$-dependence of the partial rates
$d\Gamma_{U}/dq^{2}$ (dashed), $d\Gamma_{L}/dq^{2}$ (dot-dashed) and their 
sum $d\Gamma_{U+L}/dq^{2}$ (solid)
for the $e^-$-mode (in units of $10^{-15}$~GeV$^{-1}$).
\label{fig:dUL}
}
\end{center}
\end{figure}

\begin{figure}[htb]
\begin{center}
\begin{tabular}{lr}
        \epsfig{figure=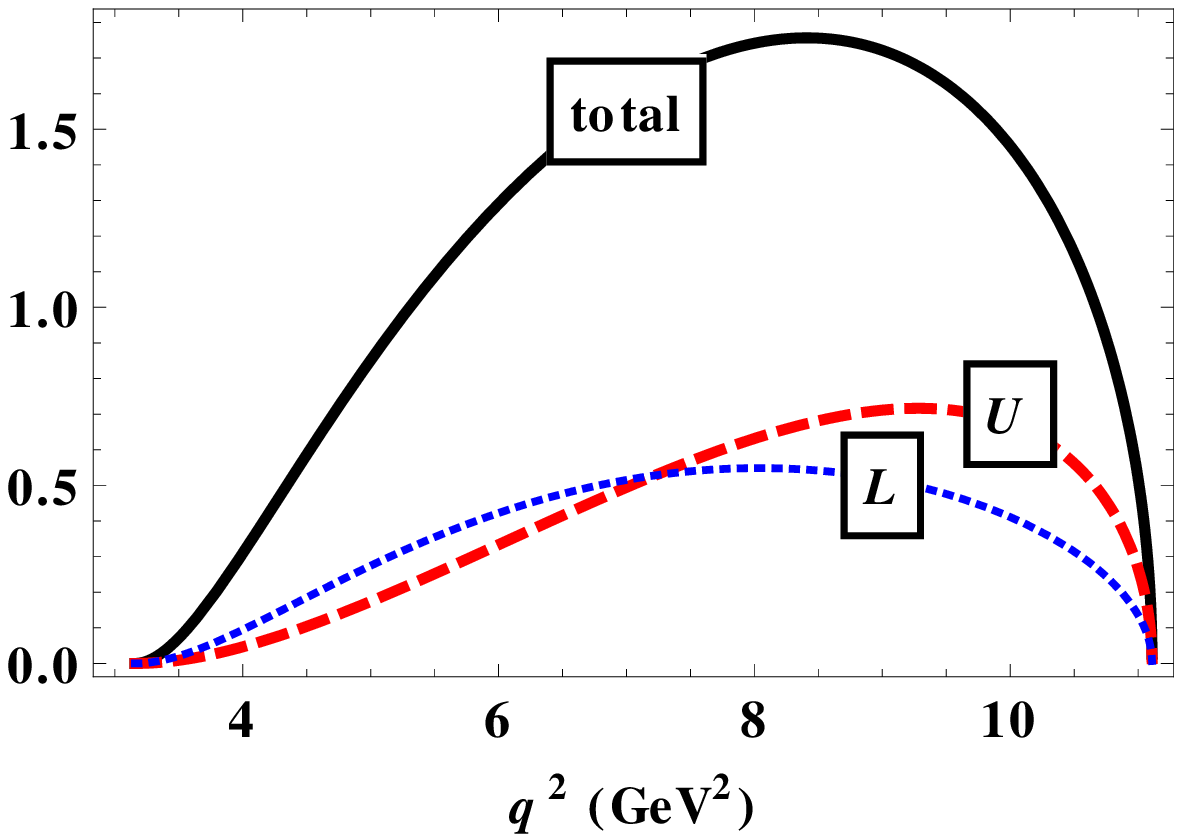,scale=0.65} 
\qquad &  \qquad  
        \epsfig{figure=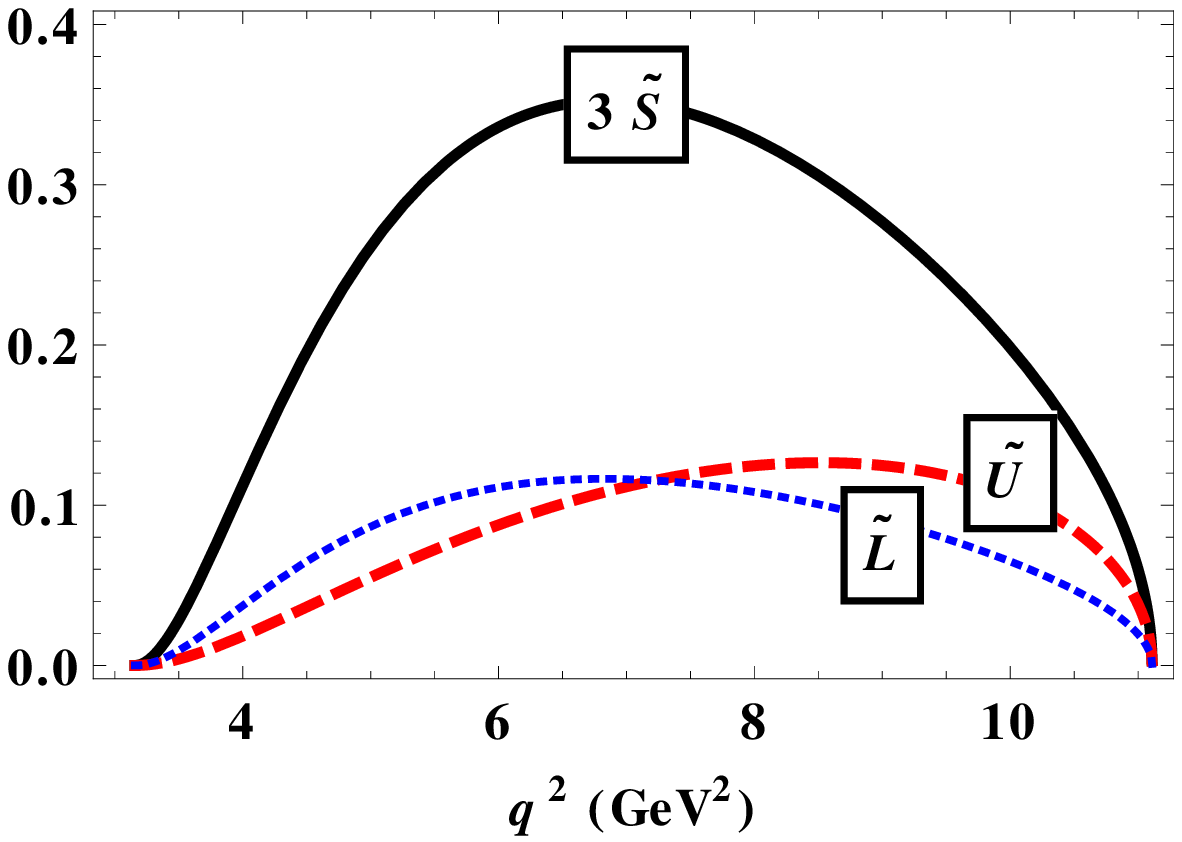,scale=0.65} 
\end{tabular}
\caption{The  $q^{2}$-dependence of the partial nonflip rates
$d\Gamma_{U,L}/dq^{2}$, and the flip rates
$d\,\widetilde \Gamma_{U,L}/d\,q^{2}$ and 
$3\,d\,\widetilde \Gamma_{S}/d\,q^{2}$ for the $\tau^-$-mode 
(in units of $10^{-15}$~GeV$^{-1}$). Also shown is the total rate 
$d\,\Gamma_{U+L}/d\,q^{2}+d\,\widetilde \Gamma_{U+L}/d\,q^{2}+
3\,d\,\widetilde \Gamma_{S}/d\,q^{2}$. 
\label{fig:dULS}
}
\end{center}
\end{figure}

In Fig.~\ref{fig:dUL}, we display the $q^{2}$ dependence of the
partial differential rates $d\Gamma_{U}/dq^{2}$, $d\Gamma_{L}/dq^{2}$ and 
the total differential rate $d\Gamma_{U+L}/dq^{2}$ for the $e$ mode.
The transverse rate dominates in the low recoil region while the longitudinal 
rate dominates in the large recoil region. The longitudinal and thereby the
total rate shows a step-like behavior close to the threshold 
$q^{2}=m^{2}_{e}$.
Figure~\ref{fig:dULS} shows the corresponding plots for the $\tau$ mode
including the partial flip rates $d\,\widetilde \Gamma_{U,L}/d\,q^{2}$ 
and $3\,d\,\widetilde \Gamma_{S}/d\,q^{2}$. We also show the total 
differential rate 
$(d\Gamma_{U+L}/dq^{2}+d\,\widetilde \Gamma_{U+L+3S}/d\,q^{2})$. 
The behavior of the
longitudinal and scalar partial flip rates can be seen to be quite close to 
the HQET result
$d\,\widetilde \Gamma_{L}/d\,q^{2}=d\,\widetilde \Gamma_{S}/d\,q^{2}$.
The helicity flip rates are smaller than the helicity nonflip rates but
contribute significantly to the total rate.

\begin{figure}[htb]
\begin{center}
\vspace*{.25cm}
\epsfig{figure=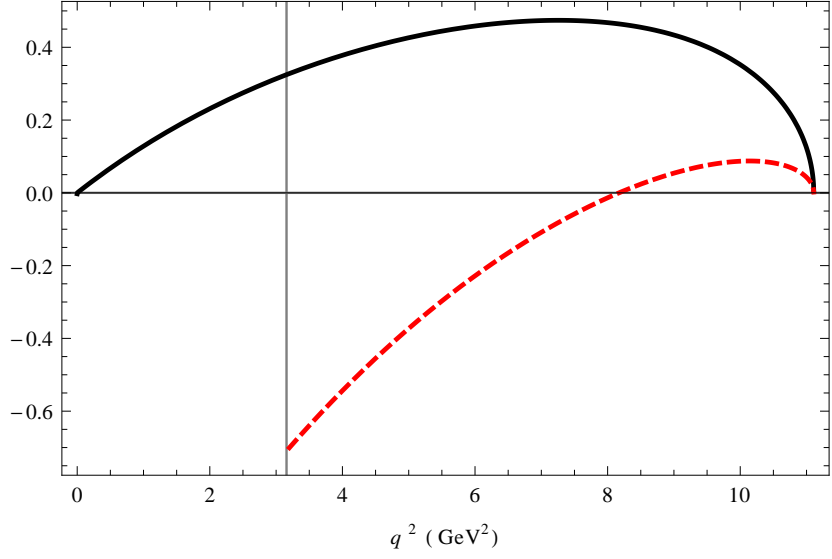,scale=.9}
\caption{The  $q^{2}$-dependence of the lepton-side forward-backward asymmetry
$A^{\ell}_{FB}(q^{2})$ for the $e^-$- (solid) and $\tau^-$-mode (dashed).
\label{fig:dFB}
}
\end{center}
\end{figure}

In Fig.~\ref{fig:dFB}, we show the $q^{2}$ dependence of the lepton-side
forward-backward asymmetry $A^{\ell}_{FB}(q^{2})$ defined in 
Eq.~(\ref{eq:FB}). 
At zero recoil, $A^{\ell}_{FB}(q^{2})$ approaches zero for both 
the $e$ and $\tau$ modes due to the zero recoil relations 
${\cal H}_{P}={\cal H}_{SL}=0$ [see Eq.~(\ref{eq:zerec})]. 
In the large recoil limit, $A^{\ell}_{FB}(e)$ goes to zero
due to the longitudinal dominance of the partial rates. The 
$q^{2}$ dependence of the forward-backward asymmetry of the two modes 
is distinctly different. While $A^{\ell}_{FB}(e)$ remains
positive in the whole $q^{2}$ range, $A^{\ell}_{FB}(\tau)$ quickly becomes
negative when moving away from zero recoil. It is interesting to observe
that $A^{\ell}_{FB}(\tau)$ goes through a zero at $q^{2}\approx 8$ GeV$^{2}$.

\begin{figure}[htb]
\begin{center}
\vspace*{.25cm}
\epsfig{figure=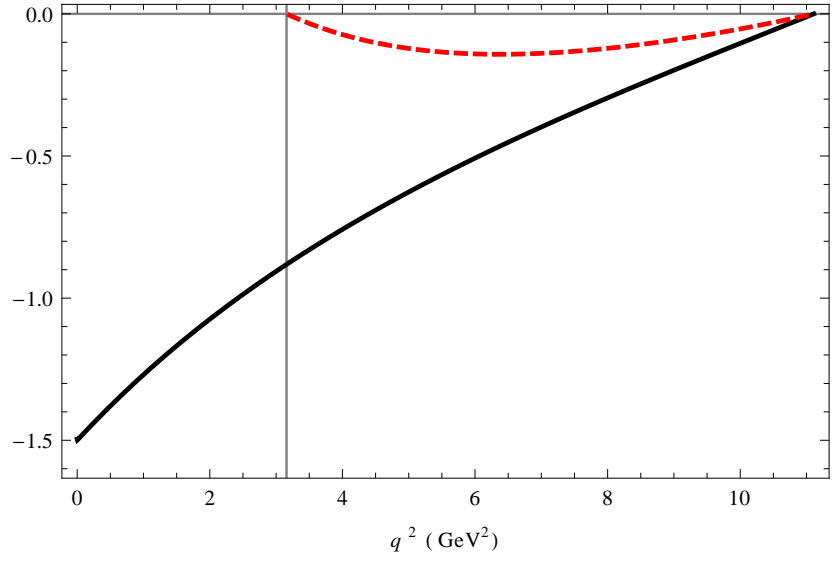,scale=.9}
\caption{The  $q^{2}$-dependence of the convexity parameter
$C_{F}(q^{2})$ for the $e^-$- (solid) and $\tau^-$-mode (dashed).
\label{fig:dCF}
}
\end{center}
\end{figure}

In Fig.~\ref{fig:dCF}, we display the $q^{2}$ dependence of the 
convexity parameter $C_{F}$ defined in Eq.~(\ref{eq:CF}). At zero recoil,  
$C_{F}$ goes to zero for both modes due to the zero recoil relation 
${\cal H}_{U}=2{\cal H}_{L}$ [see Eq.~(\ref{eq:zerec})]. 
For the $e$ mode, one finds $C_{F} \to -1,5$ at maximal recoil due to 
longitudinal dominance, while $C_{F} \to 0$ for the $\tau$ mode at maximal 
recoil $q^{2}=m^{2}_{\tau}$ due to the overall factor $(1-2\delta)$ 
in Eq.~(\ref{eq:CF}).
In the $\tau$ mode, the convexity parameter remains quite small over the whole
accessible $q^{2}$ range, implying a near straight-line behavior
of the $\cos\theta$ distribution. In the $e$ mode $C_{F}$ can become
large and negative, which implies that the $\cos\theta$ distribution is
strongly parabolic in terms of a downward open tilted parabola.  

\begin{figure}[htb]
\begin{center}
\vspace*{.25cm}
\epsfig{figure=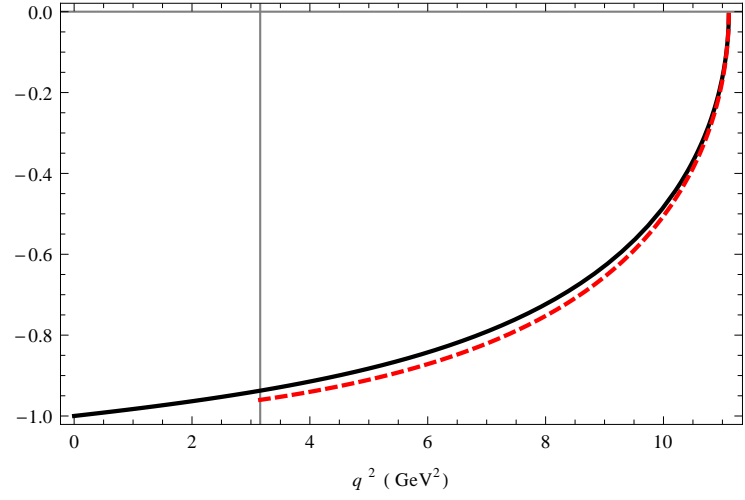,scale=.8}
\caption{The  $q^{2}$-dependence of the longitudinal polarization component
$P^h_z(q^{2})$ of the daughter baryon $\Lambda_{c}$ for the $e^-$- (solid) 
and $\tau^-$-mode (dashed).
\label{fig:dPz}
}
\end{center}
\end{figure}

\begin{figure}[htb]
\begin{center}
\vspace*{.25cm}
\epsfig{figure=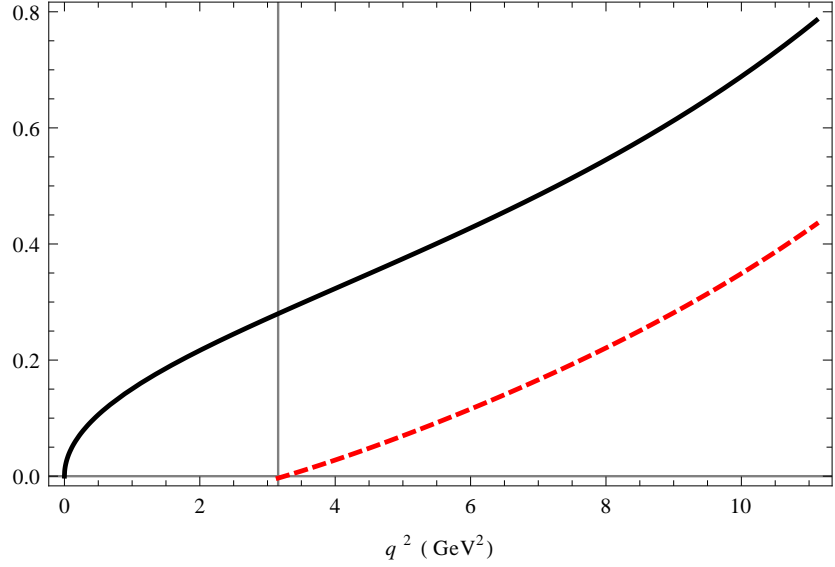,scale=.9}
\caption{The  $q^{2}$-dependence of the transverse polarization component
$P^h_x(q^{2})$ of the daughter baryon $\Lambda_{c}$ for the $e^-$- (solid) and 
$\tau^-$-mode (dashed).
\label{fig:dPx}
}
\end{center}
\end{figure}

\begin{figure}[htb]
\begin{center}
\vspace*{.25cm}
\epsfig{figure=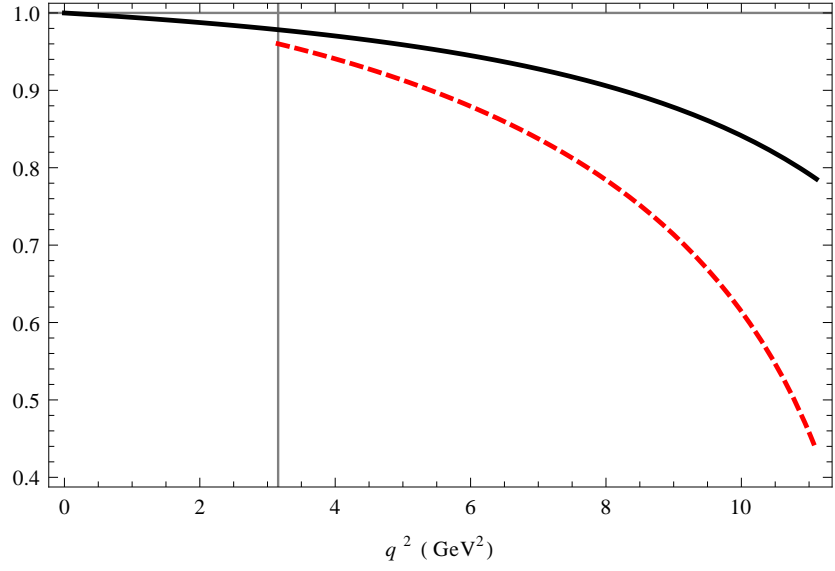,scale=.9}
\caption{The  $q^{2}$-dependence of the total $\Lambda_{c}$ polarization
$|\vec {P}\,^h|(q^{2})=\sqrt{(P^h_x)^2+(P^h_z)^2}$
for the $e^-$- (solid) and $\tau^-$-mode (dashed).
\label{fig:dPHtot}
}
\end{center}
\end{figure}

In Figs.~\ref{fig:dPz} and \ref{fig:dPx} we show the longitudinal and
transverse polarization components of the $\Lambda_{c}$ defined in 
Eq.~(\ref{eq:Pol_had}).
For the $e$ mode the longitudinal polarization goes to zero at zero recoil
since ${\cal H}_{P}={\cal H}_{L_{P}}={\cal H}_{S_{P}}=0$ at zero recoil 
[see Eq.~(\ref{eq:zerec})].
At maximal recoil, $P_{z}^{h}$ goes to zero for both modes because of the 
HQET plus $q^{2}=0$ relation Eq.~(\ref{hqet4}). The two $e$ and $\tau$ curves
almost fall on top of each other in their common $q^{2}$ range.   
The zero recoil values of the transverse polarization $P^{h}_{x}$ in 
Fig.~\ref{fig:dPx} can be understood in terms of the 
zero recoil limit of HQET [see Eq.~(\ref{hqet3})]. At zero recoil, one has  
$P_{x}\to -(\pi/4)(1-2\delta_{\ell})/(1+2\delta_{\ell})=(-0.79;-0.43)$, which
is in approximative agreement with what is shown in Fig.~\ref{fig:dPx}. 
At minimal recoil, one has
$P^{h}_{x}=0$ for both the $e$ and the $\tau$ modes. For
the $\tau$ mode, this can be understood from the approximate validity
of HQET (see the fourth equation in Eq.~(\ref{hqet3}) together with the fact
that $\delta_{\ell}=1/2$ at $q^{2}=m_{\tau}^{2}$). For the $e$ mode, one has
$P^{h}_{x}=0$ at zero recoil which follows  
from longitudinal dominance at zero recoil. The transverse 
polarization is considerably reduced going from the $e$ mode to the
$\tau$--mode. The magnitude of the $\Lambda_{c}$ polarization 
shown in Fig.~\ref{fig:dPHtot} is quite large for the $e$--mode and smaller 
for the $\tau$--mode. The maximal recoil values reflect the corresponding 
$d\,P^{h}_{x}/d\,q^{2}$ limits in Fig.~\ref{fig:dPx}, since the $z$--components
are zero at maximal recoil. 

\begin{figure}[htb]
\begin{center}
\vspace*{.25cm}
\epsfig{figure=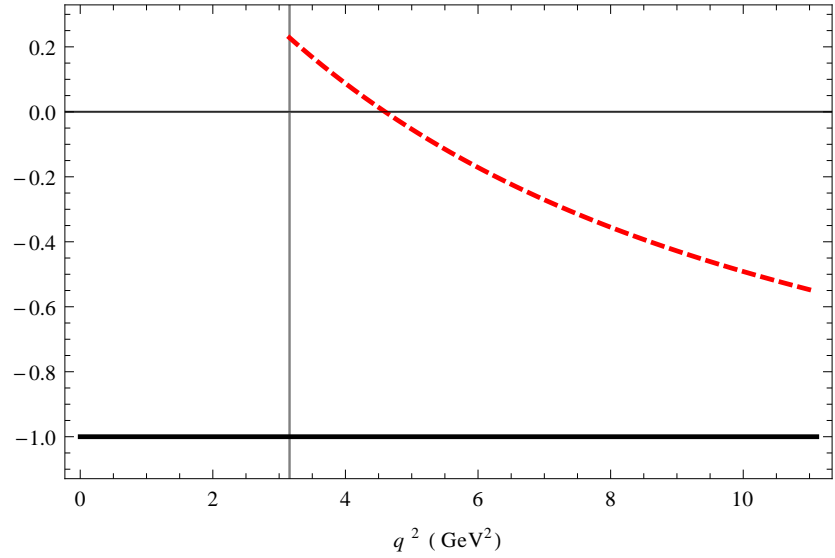,scale=.9}
\caption{The  $q^{2}$-dependence of the longitudinal polarization component
$P^\ell_z(q^{2})$ for the charged leptons $e^-$- (solid) and $\tau^-$-mode 
(dashed).
\label{fig:dPz_Lept}
}
\end{center}
\end{figure}

\begin{figure}[htb]
\begin{center}
\vspace*{.25cm}
\epsfig{figure=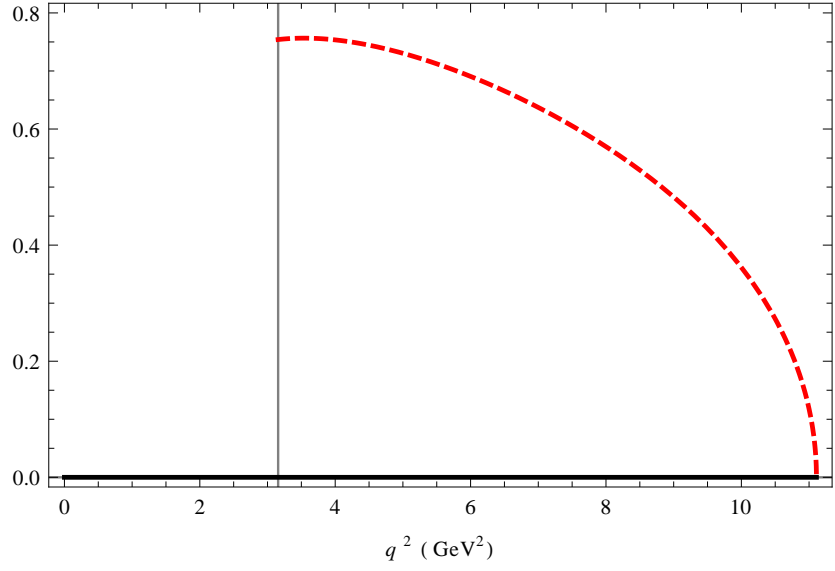,scale=.9}
\caption{The  $q^{2}$-dependence of the transverse polarization component
$P^\ell_x(q^{2})$ for the charged leptons $e^-$- (solid) and $\tau^-$-mode (dashed).
\label{fig:dPx_Lept}
}
\end{center}
\end{figure}

\begin{figure}[htb]
\begin{center}
\vspace*{.25cm}
\epsfig{figure=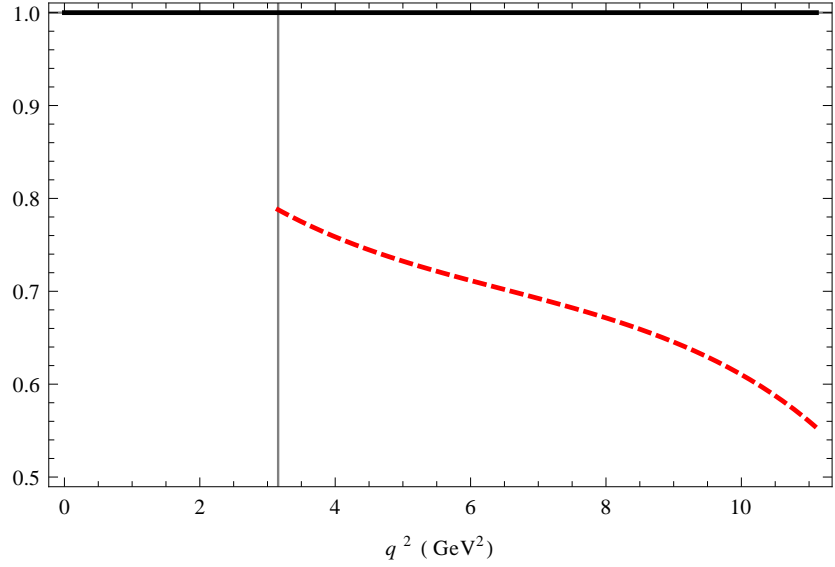,scale=.9}
\caption{The  $q^{2}$-dependence of the total lepton polarization
$|\vec{P}\,^\ell|(q^{2})=\sqrt{(P^\ell_x)^2+(P^\ell_z)^2}$
for the $e^-$- (solid) and $\tau^-$-mode (dashed).
\label{fig:dPLtot}
}
\end{center}
\end{figure}

In Figs.~\ref{fig:dPz_Lept} and~\ref{fig:dPx_Lept},  
we show the $q^{2}$ dependence of the longitudinal and
transverse polarization component of the charged lepton. In the case of
the electron, the two curves reflect the chiral limit of a massless lepton 
in which the lepton is purely left-handed. The behavior of the two
polarization components in the $\tau$ mode is distinctly different. The
longitudinal polarization is reduced from $-1$, while the transverse
polarization can become quite large towards maximal recoil. At zero recoil, 
the transverse polarization of the charged lepton $P^{\tau}_{x}$ tends to 
zero in agreement with the vanishing of 
${\cal H}_{P}$ and ${\cal H}_{SL}$ at zero recoil [see Eq.~(\ref{eq:zerec})].  
The total polarization of the lepton shown in Fig.~\ref{fig:dPLtot} is 
maximal in the $e$ mode and somewhat
reduced but still quite large in the $\tau$--mode. 

Below we present our predictions for the semileptonic branching ratios 
of the $\Lambda_b$ and $\Lambda_c$ and compare them with data. 
Consideration of both modes is sufficient 
to fix the model parameters of the $\Lambda_c$ --- 
constituent quark mass $m_c$ and the size parameter $\Lambda_{\Lambda_c}$. 
We have used the value for the $\Lambda_b$ lifetime 
from the new edition of the Particle Data Group (PDG)~\cite{Agashe:2014kda} 
$\tau_{\Lambda_b} = (1.451\pm 0.013)\cdot 10^{-12}\,$s. We have also followed
the recommendation of the PDG and have rescaled the branching ratio 
${\rm Br}(\Lambda_c^+ \to \Lambda^0 + e^+ \nu_e)$ 
upward by $30\,\%$, taking into account the new absolute measurement of 
$\mathrm{Br}(\Lambda_c^+ \to p K^- \pi^+)$ 
by the Belle Collaboration~\cite{Zupanc:2013iki}. We thus use a branching ratio
$\mathrm{Br}(\Lambda_c^+ \to \Lambda^0 + e^+ \nu_e)= (2.74\pm 0.82)\%$ 
instead of
$\mathrm{Br}(\Lambda_c^+ \to \Lambda^0 + e^+ \nu_e)= (2.1\pm 0.6)\%$ 
as listed in the 2014 edition of the PDG. 

\be
\begin{array}{ccc}
\quad \mathrm{Mode}\quad  & \quad \mathrm{Our \ results} \quad  & 
\quad \mathrm{Data} \quad  \\[1.5ex]
\hline
\Lambda_c^+ \to \Lambda^0 e^+\nu_e 
& 2.72 & 2.74 \pm 0.82 \\[1.5ex]
\hline
\Lambda_b^0 \to \Lambda_c^+ e^-\bar\nu_e 
& 6.9 & 6.5^{+3.2}_{-2.5} \\[1.5ex] 
\hline
\Lambda_b^0 \to \Lambda_c^+ \tau^-\bar\nu_\tau      
& 2.0 &  \\[1.5ex] 
\hline
\end{array}
\en 

\begin{table}[ht] 
\begin{center}
\caption{$q^{2}$ averages of helicity structure functions in units of 
$10^{-15}$~GeV. 
We do not display helicity flip results for the $e$ mode, 
because they are of order $10^{-6}-10^{-7}$ in the above units.}
\def\arraystretch{2}
\vspace*{0.5cm}
\begin{tabular}{c|ccccccccccc}
\hline
 \qquad \qquad 
&\quad $\Gamma_{U}$    \qquad   
&\quad $\Gamma_{L}$    \qquad 
&\quad $\phantom{S}$    \qquad   
&\quad $\Gamma_{LT}$   \qquad   
&\quad $\phantom{SL}$   \qquad  
&\quad $\phantom{SL}$   \qquad   
&\quad $\Gamma_{P}$    \qquad    
&\quad $\Gamma_{L_{P}}$   \qquad     
&\quad $\Gamma_{LT_{P}}$   \qquad 
&\quad $\phantom{ST_P}$ \qquad 
&\quad $\phantom{SL_P}$ \qquad 
\\
\hline
\qquad $e$\qquad  \quad   
&\quad $ 12.4 $  \qquad  
&\quad $ 19.6 $  \qquad 
&\quad $ \phantom{19.7} $  \qquad  
&\quad $-7.73 $  \qquad 
&\quad $ \phantom{2.94 }$ \qquad  
&\quad $ \phantom{18.5} $  \qquad  
&\quad $-7.61 $  \qquad 
&\quad $-18.5 $  \qquad  
&\quad $-3.50 $  \qquad  
&\quad $\phantom{-7.90} $  \qquad  
&\quad $\phantom{-19.7} $  \qquad 
\\
\qquad $\tau$ \qquad \quad   
&\quad $ 3.29 $ \qquad  
&\quad $ 2.90 $ \qquad 
&\quad $ \phantom{2.94 }$ \qquad  
&\quad $-2.06 $ \qquad 
&\quad $ \phantom{2.94 }$ \qquad  
&\quad $ \phantom{2.46} $ \qquad  
&\quad $-1.73 $ \qquad 
&\quad $-2.46 $ \qquad  
&\quad $-0.66 $ \qquad  
&\quad $\phantom{-2.10} $ \qquad  
&\quad $\phantom{-2.92} $ \qquad 
\\[2ex]
\hline
 \qquad\qquad 
&\quad $\widetilde{\Gamma}_{U}$    \qquad 
&\quad $\widetilde{\Gamma}_{L}$    \qquad 
&\quad $\widetilde{\Gamma}_{S}$    \qquad  
&\quad $\widetilde{\Gamma}_{LT}$   \qquad
&\quad $\widetilde{\Gamma}_{S_{P}}$   \qquad   
&\quad $\widetilde{\Gamma}_{SL}$   \qquad 
&\quad $\widetilde{\Gamma}_{P}$    \qquad    
&\quad $\widetilde{\Gamma}_{L_P}$   \qquad 
&\quad $\widetilde{\Gamma}_{LT_P}$   \qquad 
&\quad $\widetilde{\Gamma}_{ST_P}$ \qquad 
&\quad $\widetilde{\Gamma}_{SL_P}$ \qquad 
\\
\hline
\qquad $\tau$ \qquad \quad   
&\quad $ 0.66 $ \qquad  
&\quad $ 0.63 $ \qquad 
&\quad $ 0.64 $ \qquad  
&\quad $-0.41 $ \qquad
&\quad $-0.55 $ \qquad 
&\quad $ 0.55 $ \qquad  
&\quad $-0.37 $ \qquad 
&\quad $-0.55 $ \qquad  
&\quad $-0.14 $ \qquad  
&\quad $-0.42 $ \qquad  
&\quad $-0.64 $ \qquad 
\\[2ex]
\hline
\end{tabular}
\label{tab:bilinears-numerics}
\end{center}
\end{table}
The numbers in Table~\ref{tab:bilinears-numerics} show again that the results
of our dynamical calculation are very close to the HQET results
$\widetilde{\Gamma}_L=\widetilde{\Gamma}_S=-\widetilde{\Gamma}_{SL_{P}}$,
$\widetilde{\Gamma}_{ S_{P}}=\widetilde{\Gamma}_{L_{P}}
=-\widetilde{\Gamma}_{SL}$ and $\widetilde{\Gamma}_{ST_{P}}
=\widetilde{\Gamma}_{LT}$ 
given in terms of the helicity structure functions in Eq.~(\ref{hqet2}). 
Note, though, that the agreement with leading-order HQET 
is of effective nature only since e.g. the
size parameters of the $\Lambda_{b}$ and the $\Lambda_{c}$ are
quite different [see Eq.~(\ref{eq: size})].  

From the partial rates in Table~\ref{tab:bilinears-numerics}, 
we can compile the total rate. 
Again, we list the partial and total rates in units of 
$10^{-15}$~GeV. One has
\be
\begin{array}{cccccc|c}
   & \quad\Gamma_U\quad & \quad\Gamma_L\quad & \quad\widetilde\Gamma_{U}\quad
&\quad\widetilde\Gamma_{L}\quad&
\quad3\widetilde\Gamma_{S}\quad
& \quad\Gamma_{\rm tot}\quad \\[1.1ex]
\hline\\[-3ex]
e^-\bar\nu_e       &  12.4  &  19.6  & \cdots & \cdots & \cdots 
& 32.0   \\[1.1ex]
\tau^-\bar\nu_\tau  &  3.29  &  2.90  &0.66&0.63& 1.92  & 9.40   \\[1.1ex]
\hline
\end{array}
\label{eq:part_rates}
\en
The numbers show that the partial flip rates make up $34.2\,\%$ of the total 
rate, where the biggest contribution comes from the scalar rate with 
$20.4\,\%$. 

Next, we give the values of the integrated quantities --- 
forward-backward asymmetry $<A_{FB}>$, the asymmetry parameter $<\alpha>$, 
the convexity parameter $<C_F>$, and the hadronic  $<P^h_{x,z}>$ and 
leptonic  $<P^\ell_{x,z}>$ polarization components calculated in the model.
These can be obtained from the nonflip and flip rates collected in 
Table~{\ref{tab:bilinears-numerics}}. For example, one has 
\be
<A_{FB}^{\ell}>=\,- \frac32\, 
\frac{\Gamma_{P}+4\, \widetilde \Gamma_{SL}}{\Gamma_{tot}} \qquad
<\alpha>=<P_{z}>=\frac{\Gamma_{P}+\Gamma_{L_{P}}+ \widetilde \Gamma_{L_{P}}+
\widetilde \Gamma_{P}+3\, \widetilde \Gamma_{S_{P}}}{\Gamma_{tot}} \qquad .
\en
\be
\begin{array}{cccccccc}
       & \quad <A_{FB}^{\ell}> \quad  & 
\quad <C_F> \quad  &  \quad <P^h_z> \quad  & \quad <P^h_x>\quad 
 &  \quad <P^\ell_z> \quad  & \quad <P^\ell_x>\quad & \quad <\gamma> \qquad
\\[1.2ex]
\hline\\[-2ex]
e^-\bar\nu_e        &  
0.36  &  -0.63 & -0.82 & 0.40 & -1.00  & 0.00 & 0.31  
\\[1.2ex] 
\tau^-\bar\nu_\tau   &  
-0.077    &  -0.10 & -0.72 & 0.22 & -0.32 & 0.55 & 0.17  
\\[1.2ex] 
\hline
\end{array}
\label{eq:asymm}
\en
Note that the mean of the azimuthal asymmetry parameter $<\gamma>$ defined
in~(\ref{azipar}) is related to $<P^h_x>$ by $<\gamma>=(\pi/4)<P^h_x>$.
When calculating the $q^{2}$ averages one has to 
remember to include the $q^{2}$ dependent factor 
$(q^2-m_\ell^2)^2 |\mathbf{p_2}|/q^2$ in the numerator and denominator of the
relevant asymmetry expressions. In
most of the shown cases, the mean values change considerably when going from 
the $e$ to the $\tau$ modes including even a sign change in~$<A_{FB}^{\ell}>$. 

\section{Summary and conclusions}

Let us summarize the main results of our paper. 
We have used the helicity formalism to study the angular decay 
distribution in the semileptonic decay 
$\Lambda_b^0 \to \Lambda_c^+ + \ell^- \bar \nu_\tau$ as well as the 
corresponding cascade decay 
$\Lambda_b^0 \to \Lambda_c^+ (\to \Lambda^0 + \pi^+) + \ell^- \bar \nu_\tau$. 
Starting from the angular decay distribution, we have defined a number of 
polarization observables, for which we have provided numerical results 
using form factors obtained in 
the covariant confined quark model. All our results have been
presented for the two cases (i) the near mass zero leptons $\ell=e,\mu$ and 
(ii) the massive $\tau$ lepton $\ell=\tau$. We have demonstrated the case 
with which lepton mass effects can be included in the helicity formalism. 
The numerical results for the polarization observables
are given in the form of $q^{2}$ plots and also in averaged form, where the 
average has been taken with regard to 
the lepton--side polar angle $\cos\theta$ and also with regard to $q^{2}$. 
We find substantial lepton mass effects in most of the polarization 
observables. We have also discussed the decay  
$\bar\Lambda_b^0 \to \bar\Lambda_c^- 
(\to \bar\Lambda^0 + \pi^{-}) + \ell^+ \nu_\ell$,
the angular decay distribution of which is identical to that of the decay 
$\Lambda_b^0 \to \Lambda_c^+ (\Lambda^0 + \pi^{+}) + \ell^- \bar\nu_\ell$ 
after implementing the appropriate changes. In our analysis we have assumed 
that the parent baryon $\Lambda_b$ is unpolarized.  
The case of polarized $\Lambda_b$ decays would bring in a number of
new polarization observables as discussed 
in corresponding calculations 
in Refs.~\cite{Gutsche:2013oea,Bialas:1992ny,Kadeer:2005aq}.

In our analysis we have remained within the SM. The helicity formalism 
for the decays presented in this paper easily allows one to include 
effects beyond the BSM. 
A generic charged Higgs contribution would populate 
the scalar  helicity amplitudes $H_{+\frac12 t}^{V/A}$ and would thus
affect the helicity structure functions ${\cal H}_{S},\,{\cal H}_{SL}\,
{\cal H}_{SP},\,$ and 
${\cal H}_{ST_{P}}$ through their interference with the SM contributions. 
Tensor currents contribute only to the spin-1 piece of the helicity
amplitudes $H_{+\frac12 +1}^{V/A}$ and
$H_{+\frac12 0}^{V/A}$ . They would populate all helicity structure 
functions when interfering with the SM amplitudes except for ${\cal H}_{S}$.
Depending on the chiral structure of the lepton-side effective currents, the  
pattern of helicity flip and nonflip contributions to the helicity structure
functions would be redistributed. 
It would be interesting to find out how many of the BSM proposals 
that have been introduced in the meson sector to explain the 
tensions with the SM results would have a measurable effect in
the baryon sector.

\begin{acknowledgments}

This work was supported by the Tomsk State University Competitiveness 
Improvement Program.  
M.A.I.\ acknowledges the support from the Mainz Institute for Theoretical 
Physics (MITP). M.A.I. and J.G.K. thank the Heisenberg-Landau Grant for
support.  

\end{acknowledgments}

\appendix
\section{Unpolarized and polarized lepton tensor} 

In this appendix we determine the helicity components of the unpolarized and 
polarized lepton tensor in the $\Lambda_{b}$ decay system ($z$ axis
along the momentum of the off-shell $W$) using covariant methods.  
As shown in Eq.~(\ref{rot2}) the unpolarized and polarized helicity 
components in the $\Lambda_{b}$ decay system can also be 
obtained by a rotation of the helicity components of the lepton 
tensor in the $W$--decay system ($z$ axis along $\ell^{\mp}$).
This provides for a check of the covariant calculation. 
In the following, the helicity components of the lepton tensor are always given 
for both the $(\ell^-\bar\nu_\ell)$ and the $(\ell^+\nu_\ell)$ configurations. 
The latter configuration is needed for the decay 
$\bar\Lambda_b^0 \to \bar\Lambda_c^- + \ell^+ \nu_\ell$ 
or for the decay $\Lambda_c^+ \to \Lambda^0 + \ell^+ \nu_\ell$.

\subsection{Unpolarized lepton}

The unpolarized lepton tensor for the process 
       $W^-_{\rm off-shell}\to \ell^-\bar \nu_\ell$ 
$\left( W^+_{\rm off-shell}\to \ell^+ \nu_\ell \right)$  is calculated as

\bea
L^{\mu\nu} &=& \left\{\begin{array}{lr}
\Tr\,\Big[ ( \slp_\ell + m_\ell) \gamma^\mu(1-\gamma_5)
           \slp_{\nu_\ell} \gamma^\nu (1-\gamma_5)\Big]
\qquad\text{for}\qquad  W^-_{\rm off-shell}\to \ell^-\bar\nu_\ell 
\\[1.2ex]
\Tr\,\Big[ (\slp_\ell - m_\ell) \gamma^\nu(1-\gamma_5)
             \slp_{\nu_\ell} \gamma^\mu(1-\gamma_5)\Big]
\qquad\text{for}\qquad  W^+_{\rm off-shell}\to \ell^+ \nu_\ell 
                   \end{array}\right.
\nn[1.2ex]
&=&
8\, \left( 
  p_\ell^\mu p_{\nu_\ell}^\nu  + p_\ell^\nu p_{\nu_\ell}^\mu 
- p_{\ell}\cdot p_{\nu_\ell}\,g^{\mu\nu}  \pm  
i \varepsilon^{\mu \nu \alpha \beta} p_{\ell\,\alpha} p_{\nu_\ell\,\beta}
\right)
\label{eq:lept_tensor}
\ena
where the upper/lower sign 
refers to the two $(\ell^-\bar\nu_\ell)/(\ell^+\nu_\ell)$ configurations.
The sign change can be seen to result from the p.v. part of the lepton 
tensors.
 
\begin{figure}[htb]
\begin{center}
\epsfig{figure=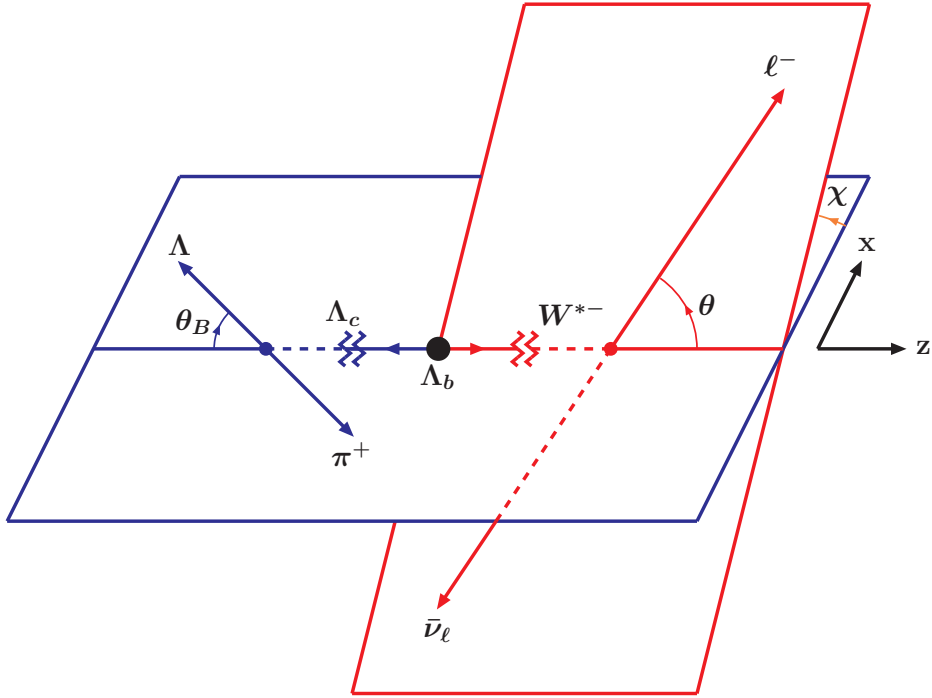,scale=.65}
\caption{Definition of the polar angles $\theta$, $\theta_B$ and the
azimuthal angle $\chi$.} 
\label{fig:angles}
\end{center}
\end{figure}

We shall evaluate the lepton tensor components in the $W_{\rm off-shell}$ 
rest frame with the $z$--axis along the original momentum direction of the
$W_{\rm off-shell}$ (see Fig.~\ref{fig:angles}). The polarization 
four-vectors in the $W_{\rm off-shell}$ rest frame with 
helicities $\lambda_{W}=t,\pm,0$ are given by
\be
\epsilon^\mu(t)=
(\,1\,,\,\,0\,,\,\,0\,,\,\,0\,)
\qquad
\epsilon^\mu(\pm) = 
\frac{1}{\sqrt{2}} (\,0\,,\,\,\mp 1\,,\,\,-i\,,\,\,0\,) 
\qquad
\epsilon^\mu(0) =
(\,0\,,\,\,0\,,\,\,0\,,\,\,1\,)\,.
\en
The momentum vectors $p_\ell^\mu$ and $p_{\nu_\ell}$ in the same system 
read (see Fig.~\ref{fig:angles}) 
\be\label{EqB1}  
p^\mu_\ell = (E_\ell; |{\bf p_\ell}|\sin\theta\cos\chi, 
|{\bf p_\ell}|\sin\theta\sin\chi, 
|{\bf p_\ell}|\cos\theta)\,, \quad 
p^\mu_{\nu_\ell} = |{\bf p_\ell}| (1; - \sin\theta\cos\chi, 
- \sin\theta\sin\chi , 
-\cos\theta)\,, 
\en
where $E_\ell = (q^2 + m_\ell^2)/(2\sqrt{q^2})$ and 
$|{\bf p_\ell}| = (q^2 - m_\ell^2)/(2\sqrt{q^2})=\sqrt{q^{2}}v/2$ are 
the energy and momenta of the charged lepton.
The helicity components $L_{\lambda_{W}\lambda'_{W}}(\theta,\chi)$ can then 
be obtained from the contraction
\be
L_{\lambda_{W}\lambda'_{W}}(\theta,\chi)=L^{\mu\nu}(\theta,\chi)\epsilon_{\mu}(\lambda_{W})
\epsilon^{\ast}_{\nu}(\lambda'_{W})
\label{contr}
\en
One obtains (the rows and columns of the matrix are ordered in the sequence 
$(t,1,0,-1)$)
\bea 
\label{lt1}
(2q^2v)^{-1} L_{\lambda_{W}\lambda'_{W}}(\theta,\chi) &=& 
\left( \begin{array}{cccc} 
0 & 0 & 0 & 0 \\
0 & (1\mp\cos\theta)^2 & \mp \frac{2}{\sqrt{2}} (1\mp\cos\theta) \sin\theta 
e^{i\chi} & \sin^2\theta e^{2i\chi} \\
0 & \mp \frac{2}{\sqrt{2}} (1\mp\cos\theta) \sin\theta 
e^{-i\chi} & 2\sin^2\theta 
& \mp \frac{2}{\sqrt{2}} (1\pm\cos\theta) \sin\theta 
e^{i\chi} \\
0 & \sin^2\theta e^{-2i\chi} 
& \mp \frac{2}{\sqrt{2}} (1\pm\cos\theta) \sin\theta 
e^{-i\chi} & (1\pm\cos\theta)^2 \\
\end{array} \right)\nonumber\\
&+&\delta_\ell 
\left( \begin{array}{cccc} 
4 & - \frac{4}{\sqrt{2}} \sin\theta e^{i\chi} & 4 \cos\theta &
 \frac{4}{\sqrt{2}} \sin\theta e^{i\chi} \\
- \frac{4}{\sqrt{2}} \sin\theta e^{-i\chi} & 
2 \sin^2\theta & 
- \frac{2}{\sqrt{2}} \sin 2\theta e^{i\chi} & 
- 2 \sin^2\theta e^{2i\chi} \\
4\cos\theta & 
- \frac{2}{\sqrt{2}} \sin 2\theta e^{-i\chi} & 
4\cos^2\theta & 
\frac{2}{\sqrt{2}} \sin 2\theta e^{i\chi} \\
\frac{4}{\sqrt{2}} \sin\theta e^{-i\chi} & 
- 2 \sin^2\theta e^{-2i\chi} & 
\frac{2}{\sqrt{2}} \sin 2\theta e^{-i\chi} & 
2\sin^2\theta \\
\end{array} \right)
\ena 
The components $L_{1-1}(\theta,\chi)$ and $L_{-11}(\theta,\chi)$ are not 
probed in unpolarized
$\Lambda_b$ decays because of the condition
$|\lambda_W - \lambda'_W| \le 1$ discussed after Eq.~(\ref{set}). These 
two components are, however, probed in polarized $\Lambda_b$ decays.

When one integrates Eq.~(\ref{lt1}) over the azimuthal angle $\chi$ one 
obtains the quasi-diagonal form
\be
\label{lt2}
\frac{1}{2\pi}(2q^2v)^{-1} L_{\lambda_{W}\lambda'_{W}}(\theta) = 
\left( \begin{array}{cccc} 
0 & 0 & 0 & 0 \\
0 & (1\mp\cos\theta)^2 &0 & 0 \\
0 & 0 & 2\sin^2\theta 
& 0 \\
0 & 0 &0 & (1\pm\cos\theta)^2 \\
\end{array} \right)
+\delta_\ell 
\left( \begin{array}{cccc} 
4 & 0 & 4 \cos\theta &
 0 \\
0 & 
2 \sin^2\theta & 
 & 0 \\
4\cos\theta & 
0 & 
4\cos^2\theta & 
0 \\
0 & 0 & 
0 & 
2\sin^2\theta \\
\end{array} \right)
\en
where $L_{\lambda_{W}\lambda'_{W}}(\theta)=
\int d\chi\,  L_{\lambda_{W}\lambda'_{W}}(\theta,\chi)$.

Finally, integrating Eq.~(\ref{lt2}) over $\cos\theta$ one obtains the diagonal
form
\be
\frac{1}{2\pi}(2q^2v)^{-1} L_{\lambda_{W}\lambda'_{W}} = \frac83 
\left( \begin{array}{cccc} 
3\delta_{\ell} & 0 & 0 & 0 \\
0 &  (1+\delta_{\ell}) &0 & 0 \\
0 & 0 &  (1+\delta_{\ell})
& 0 \\
0 & 0 &0 &  (1+\delta_{\ell}) \\
\end{array} \right)
\en
where $L_{\lambda_{W}\lambda'_{W}}=
\int d\cos\theta\,  L_{\lambda_{W}\lambda'_{W}}(\theta)$.

Next we calculate the lepton helicity tensor 
$\hat L_{\hat\lambda_{W}\hat\lambda'_{W}}$ in the $W$--decay system
with the $z$-axis along the charged lepton. In order to set it apart from the
lepton helicity tensor in the $\Lambda_{b}$-decay system, we use a hat
notation such that $\hat\lambda_{W}=\lambda_{\ell}-\lambda_{\nu}$. 
The components of $\hat L_{\hat\lambda_{W}\hat\lambda'_{W}}$ can be 
calculated as in~(\ref{contr}), but now using
\be\label{EqB2}  
p^\mu_\ell = (E_\ell;0, 
0, |{\bf p_\ell}|)\,, \quad 
p^\mu_{\nu_\ell} = |{\bf p_\ell}| (1; 0, 0, -1)\,. 
\en
The nonvanishing components read
\be
(2q^2v)^{-1} \hat{L}_{\mp\mp} = 4 \qquad 
(2q^2v)^{-1} \hat{L}_{00} = 
(2q^2v)^{-1} \hat{L}_{t0} = 
(2q^2v)^{-1} \hat{L}_{0t} = 
(2q^2v)^{-1} \hat{L}_{tt} = 4\delta_\ell \,. 
\label{Lhatunpol}
\en 

The helicity components $L_{\lambda_{W}\lambda'_{W}}(\theta,\chi)$ in the
$\Lambda_{b}$-decay system can be obtained 
from the $\hat L_{\hat\lambda_{W}\hat\lambda'_{W}}$ in the $W$--decay system 
by an appropriate rotation. One has 
\be
L_{\lambda_{W}\lambda'_{W}}(\theta,\chi)=
\sum_{\hat \lambda_{W},\hat \lambda'_{W}=t,\mp1,0}
d^{J}_{\lambda_{W},\hat \lambda_{W}}(\theta)\,
d^{J'}_{\lambda'_{W},\hat \lambda'_{W}}(\theta)\,
e^{i(\lambda_{W}-\lambda'_{W})\chi}\,
\hat L_{\hat\lambda_{W}\hat\lambda'_{W}} .
\en

\subsection{Longitudinal polarization $P^{\ell}_z$}
Next we calculate the longitudinal polarization component $P_{z}$ of 
the 
charged lepton. The relevant lepton tensor  
can be obtained from the unpolarized tensor 
by the substitution $p_\ell^\mu \to \mp m_\ell s_z^\mu$:   
\be 
L^{\mu\nu}(s_{z}) =  \mp 8 m_\ell \biggl(s_z^\mu p_{\nu_\ell}^\nu 
                             + s_z^\nu p_{\nu_\ell}^\mu 
              - s_z \cdot p_{\nu_\ell} g^{\mu\nu} 
              \pm  i \varepsilon^{\mu \nu \alpha \beta} 
              s_{z\,\alpha} p_{\nu_\ell\,\beta} \biggr) \, .
\label{eq:long_Pz} 
\en  
At present, we shall not consider the $\chi$ dependence of the longitudinal 
polarization component $P_{z}$. Without loss of generality, one can then set
$\chi= 0$ in the representation of the momentum $p^\mu_{\nu_\ell}$ in 
Eq.~(\ref{EqB1}) but must compensate by the factor $2\pi$ from the trivial 
$\chi$ integration.
The four-component spin vector $s_z^\mu$ reads ($\chi= 0$)
\be 
s_z^\mu = \frac{1}{m_\ell} (|{\bf p}_\ell|; 
E_\ell\sin\theta, 0, E_\ell\cos\theta)
\label{sz}
\en 
which satisfies the conditions $s_{z, \mu} s_z^\mu = -1$ and 
$s_{z, \mu} \cdot p_\ell^\mu = 0$.  

The components $L_{\lambda_{W}\lambda'_{W}}(s_z;\theta)$  can be 
calculated to be  
\bea
\frac{1}{2\pi}\,(2q^2v)^{-1} L_{tt}(s_z;\theta) = 
\pm 4 \delta_{\ell}\qquad
&\to&\pm 8\delta_\ell\,,\nonumber\\
\frac{1}{2\pi}\,(2q^2v)^{-1} L_{t0}(s_z;\theta) = 
\frac{1}{2\pi}\,(2q^2v)^{-1} L_{0t}(s_z;\theta) 
= \pm4 \delta_{\ell}\cos\theta\qquad&\to& 0\,,\nonumber\\
\frac{1}{2\pi}\,(2q^2v)^{-1} L_{++}(s_{z};\theta) = 
-(1\mp\cos\theta)^2 + 2\delta_\ell\sin^2\theta 
\qquad&\to& \mp\frac{8}{3} (1-\delta_\ell)\,,\nonumber\\
\frac{1}{2\pi}\,(2q^2v)^{-1} L_{00}(s_z;\theta) = 
-2\sin^2\theta + 4\delta_\ell\cos^2\theta 
\qquad&\to&\mp \frac{8}{3} (1-\delta_\ell)\,,\nonumber\\
\frac{1}{2\pi}\,(2q^2v)^{-1} L_{--}(s_z;\theta) = 
-(1\pm\cos\theta)^2 + 2\delta_\ell\sin^2\theta 
\qquad&\to&\mp \frac{8}{3} (1-\delta_\ell) \,.
\ena 
The nonvanisihing helicity components 
$\hat L_{\hat\lambda_{W}\hat\lambda'_{W}}$ in the $W$--decay helicity system 
are obtained by setting $\theta=0$ in (\ref{sz}) such that 
$s_z^\mu =  (|{\bf p}_\ell|; 0, 0, E_\ell)/{m_\ell} $ and taking 
$p^\mu_{\nu_\ell}$ as in~(\ref{EqB2}).
One obtains
\be  
(2q^2v)^{-1} \hat{L}_{\mp\mp}(s_{z}) =\mp 4\,, \quad 
(2q^2v)^{-1} \hat{L}_{00}(s_{z}) = 
(2q^2v)^{-1} \hat{L}_{t0}(s_{z}) = 
(2q^2v)^{-1} \hat{L}_{0t}(s_{z}) = 
(2q^2v)^{-1} \hat{L}_{tt}(s_{z}) = 
\pm4\delta_\ell \,.
\label{Lhatpolz}
\en 
The helicity components $L_{\lambda_{W}\lambda'_{W}}(s_{z};\theta)$ in the
$\Lambda_{b}$-decay system can again be obtained by rotation:  
\be 
\sum\limits_{\hat \lambda_{W},\hat \lambda'_{W}}
L_{\lambda_{W}\lambda'_{W}}(s_{z},\theta)=
\sum\limits_{\hat \lambda_{W},\hat \lambda'_{W}}
d^{J}_{ \lambda_{W}
\hat \lambda_{W}}(\theta)\,
d^{J'}_{ \lambda'_{W}\hat \lambda'_{W}}(\theta)\,
\hat{L}_{\hat \lambda_{W}\hat \lambda'_{W}}(s_{z}) .
\en 
 
\subsection{Transverse polarization $P^{\ell}_x$}

The lepton tensor relevant for the transverse polarization $P_x$ of the 
charged lepton 
is obtained from the unpolarized expression 
by the substitution $p_\ell^\mu \to \mp m_\ell s_x^\mu$. Again we 
set $\chi=0$ without loss of generality. The polarized lepton
tensor reads   
\be 
L^{\mu\nu}(s_{x}) =  \mp8 m_\ell \biggl(s_x^\mu p_{\nu_\ell}^\nu 
                             + s_x^\nu p_{\nu_\ell}^\mu 
              - s_x \cdot p_{\nu_\ell} g^{\mu\nu} 
              \pm  i \varepsilon^{\mu \nu \alpha \beta} 
              s_{x\,\alpha} p_{\nu_\ell\,\beta}\biggr) \,,
\label{eq:long_Px} 
\en  
where $s_x^\mu$ is the four-component spin vector 
\be 
s_x^\mu = (0; 
\cos\theta, 0, -\sin\theta)\,. 
\label{xspin}
\en 
obeying the conditions $s_{x, \mu} s_x^\mu = -1$ and 
$s_{x, \mu} \cdot p_\ell^\mu = 0$. Just as before, the relevant components
of the polarized lepton tensor are obtained by contraction with the
relevant polarization four-vectors. One obtains
\bea
\frac{1}{2\pi}\,(2q^2v)^{-1} L_{tt}(s_x;\theta) = 0      
\qquad &\to&0\,,\nonumber\\
\frac{1}{2\pi}\,(2q^2v)^{-1} L_{t0}(s_x;\theta) 
= (2 q^2 v)^{-1} L_{0t}(s_x;\theta) 
= \mp \sqrt{\delta_\ell} \frac{4}{\sqrt{2}} \sin\theta
\qquad&\to& 
\mp 2\pi\sqrt{\delta_{\ell}/2}\,,\nonumber\\
\frac{1}{2\pi}\,(2q^2v)^{-1} L_{++}(s_x;\theta) = 
- \sqrt{\delta_\ell} \frac{4}{\sqrt{2}} \sin\theta
(1\mp\cos\theta) 
\qquad&\to& - 2\pi\sqrt{\delta_{\ell}/2}\,,\nonumber\\
\frac{1}{2\pi}\,(2q^2v)^{-1} L_{00}(s_x;\theta) = 
\mp \sqrt{\delta_\ell} \frac{8}{\sqrt{2}} \sin\theta\cos\theta  
\qquad&\to& 0\,,\nonumber\\
\frac{1}{2\pi}\,(2q^2v)^{-1} L_{--}(s_x;\theta) =  
\sqrt{\delta_\ell} \frac{4}{\sqrt{2}} \sin\theta (1\pm\cos\theta) 
\qquad&\to& + 2\pi\sqrt{\delta_{\ell}/2}\,.
\ena 
The nonvanishing helicity components in the
$W$-decay system are obtained by setting $\theta=0$ in~(\ref{xspin}) such that
$s_x^\mu = (0; 1, 0, 0)$. One obtains 
\be 
(2q^2v^2)^{-1} (\hat{L}_{t\mp}) =(2q^2v^2)^{-1} (\hat{L}_{\mp t})= 
(2q^2v^2)^{-1} \hat{L}_{0\mp} =(2q^2v^2)^{-1} (\hat{L}_{\mp 0})=  
 4\sqrt{\delta_\ell} \, .
\label{Lhatpolx}
\en 
As before the relevant helicity components in the production system are
obtained by a rotation via
\be
\sum\limits_{\hat \lambda_{W},\hat \lambda'_{W}} 
L_{\lambda_{W}\lambda'_{W}}(s_{x},\theta)=
\sum\limits_{\hat \lambda_{W},\hat \lambda'_{W}} 
d^{J}_{ \lambda_{W}
\hat \lambda_{W}}(\theta)\,
d^{J'}_{ \lambda'_{W}\hat \lambda'_{W}}(\theta)\,
\hat{L}_{\hat \lambda_{W}\hat \lambda'_{W}}(s_{x}) .
\en

\section{Lepton-side helicity amplitudes}
In this appendix we calculate the lepton--side helicity amplitudes 
appearing in Eqs.~(\ref{Lhat}), (\ref{rot2}) and (\ref{eq:fourfold2}). 
The lepton--side helicity amplitudes $h_{\lambda_\ell,1/2}$ for 
the process 
$W^{-}_{\rm off-shell} \to \ell^{-}\bar \nu_{\ell}$ are defined by 
$(\hat \lambda_W=\lambda_\ell-1/2)$ 
\be
h_{\hat{\lambda}_{W};\lambda_{\ell^{-}},1/2}(J)=
\bar{u}_{\ell^{-}}(\lambda_{\ell})\,\gamma_{\mu}(1-\gamma_{5})
\,v_{\bar \nu_{\ell}}(1/2)
\,\epsilon^\mu(\hat{\lambda}_{W})\, .
\label{eq:defhl-}
\en
As before, we use the hat notation for the helicities in the 
$W$--decay system. 
We evaluate the helicity amplitudes in the $(\ell^{-}\bar \nu_{\ell})$ 
center-of-mass (CM) system,  
with $\ell^{-}$ defining the $z$--direction such that
$\hat{\lambda}_{W}=\lambda_{\ell}-1/2$.
The label $(J)$ takes the values 
$(J=0)$ with $\lambda_{W}=0\,(:=\,t)$ and $(J=1)$ with $\lambda_{W}=\pm1,0$ 
for the lepton current $J^{\mu}_{W}$, respectively. The relevant spinors 
are given by~\cite{ab66}
\be
\label{spinor}
\bar{u}_2(\pm{\textstyle \frac{1}{2}},p_{\ell})= \sqrt{E_{\ell}+m_\ell}
\left( \chi_\pm^\dagger,
\frac{\mp|{\bf p_\ell}|}{E_\ell+m_\ell}\chi_\pm^\dagger \right)\,, \qquad
v_{\bar{\nu}}({\textstyle \frac{1}{2}},p_{\bar\nu_\ell}) = \sqrt{E_\nu}
\left(\begin{array}{c}\chi_+ \\-\chi_+  \end{array}\right) \, , 
\en
where $\chi_+=\binom{1}{0}$ and $\chi_-=\binom{0}{1}$ 
are the usual Pauli two-spinors.
The lepton--side helicity amplitudes can be calculated to be 
\bea
h_{\,t\,;+1/2\,+1/2}(J=0) &=&  2\, \sqrt{v}\, m_\ell \,,
\nn
h_{\,0\,;+1/2\,+1/2}(J=1) &=&  2\, \sqrt{v}\,  m_\ell \,,
\nn
h_{ -1\,;-1/2\,+1/2}(J=1) &=&  2\sqrt{2}\, \sqrt{q^{2}}\, \sqrt{v}\,,   
\label{eq:lhelamp-}
\ena
where, as before, $v=1-m_{\ell}^2/q^2$ 
is the lepton velocity in the $(\ell^-\bar \nu_{\ell})$ CM frame. 
On squaring the lepton-side $V-A$ helicity amplitudes one finds 
\bea
    |h_{\,t\,;+1/2\,+1/2}|^2 
&=& |h_{\,0\,;+1/2\,+1/2}|^2
 = 4m_\ell^2 v = 8q^2 v\,\delta_\ell\,,  
\nn
 h_{\,t\,;+1/2\,+1/2}\,h_{\,0\,;+1/2\,+1/2}
&=&
4m_\ell^2 v = 8q^2 v\,\delta_\ell\,,
\nn
|h_{-1\,;-1/2\,+1/2}|^{2} &=& 8q^2 v\,.
\nonumber
\ena
The helicity flip contributions can be seen to be suppressed relative to the 
helicity nonflip contributions by the factor 
$\delta_\ell=m_{\ell}^{2}/2q^{2}$, i.e.
$|h_{hf}|^{2}/|h_{nf}|^{2}=\delta_\ell$.

One can go through the same exercise for 
$W^{+}_{\rm off-shell} \to \ell^{+} \nu_{\ell}$ with 
the $z$--direction defined by $\ell^{+}$, i.e. 
$(\hat{\lambda}_{W}=\lambda_{\ell}+1/2)$. The helicity amplitudes are defined
by
\be h_{\hat{\lambda}_{W};\lambda_{\ell^{+}},-1/2}(J)=
\bar{u}_{\nu_{\ell}}(-1/2)\,\gamma_{\mu}(1-\gamma_{5})
\,v_{{\ell^{+}}}(\lambda_{\ell^+})
\,\epsilon^\mu(\hat{\lambda}_W)\,,
\label{eq:defhl+}
\en
which can be evaluated with the explicit forms of the spinors
\be
\label{spinor2}
\bar{u}_{\nu_{\ell}}(-{\textstyle \frac{1}{2}},p_{\nu_{\ell}})= 
-\,\sqrt{E_{\nu}}
\left( \chi_+^\dagger\,,\chi_+^\dagger \right) \qquad
v_{\ell^+}(\pm{\textstyle \frac{1}{2}},p_{\ell}) = \sqrt{E_\ell+m_{\ell}}
\left(\begin{array}{c}\frac{-|{\bf p_\ell}|}{E_\ell+m_\ell}\chi_\mp \\
\pm\chi_\mp \end{array}\right) \, , 
\en
and one finds 
\bea
h_{\,t\,;-1/2\,-1/2}\,(J=0) &=&  - 2\, \sqrt{v}\, m_\ell \,,
\nn
h_{\,0\,;-1/2\,-1/2}\,(J=1) &=&  - 2\, \sqrt{v}\, m_\ell \,,
\nn
h_{ +1\,;+1/2\,-1/2}\,(J=1) &=&  - 2\sqrt{2}\, \sqrt{q^{2}}\, \sqrt{v}  .
\label{eq:lhelamp+}
\ena
The lepton tensor helicity components for the two cases are given by 
(we assume the lepton-side helicity amplitudes to be relatively real)
\bea
\text{unpolarized\, case:}\qquad\qquad 
\hat L_{\hat\lambda_{W}\hat\lambda'_{W}}
&=&
\sum_{\lambda_{\ell}}\,h_{\hat\lambda_{W};\lambda_{\ell}\,\pm1/2}
h_{\hat\lambda'_{W};\lambda_{\ell}\,\pm1/2} 
\nn
\text{longitudinal polarization:}\qquad 
\hat L_{\hat\lambda_{W}\hat\lambda'_{W}}(s_{z})
&=&
|h_{\hat\lambda_{W};1/2\,\pm1/2}|^{2}-
|h_{\hat\lambda'_{W};-1/2\,\pm1/2}|^{2} 
\nn
\text{transverse polarization:}\qquad \hat 
L_{\hat\lambda_{W}\hat\lambda'_{W}}(s_{x})
&=&2\,h_{\hat\lambda_{W};1/2\,\pm1/2}h_{\hat\lambda'_{W};-1/2\,\pm1/2}
\ena
Inserting the lepton-side helicity amplitudes, one reproduces the results 
of Eqs.~(\ref{Lhatunpol}), (\ref{Lhatpolz}) and (\ref{Lhatpolx}).

\vspace*{4cm}

\newpage 

\clearpage 

\vspace*{2cm}

\ed